\documentstyle[onecolumn,epsf]{mn}

\newif\ifAMStwofonts
\ifoldfss
  \ifCUPmtlplainloaded \else
    \NewTextAlphabet{textbfit} {cmbxti10} {}
    \NewTextAlphabet{textbfss} {cmssbx10} {}
    \NewMathAlphabet{mathbfit} {cmbxti10} {} % for math mode
    \NewMathAlphabet{mathbfss} {cmssbx10} {} %  "   "    "
  \fi
  \ifAMStwofonts
    \ifCUPmtlplainloaded \else
      \NewSymbolFont{upmath} {eurm10}
      \NewSymbolFont{AMSa} {msam10}
      \NewMathSymbol{\upi}     {0}{upmath}{19}
      \NewMathSymbol{\umu}     {0}{upmath}{16}
      \NewMathSymbol{\upartial}{0}{upmath}{40}
      \NewMathSymbol{\leqslant}{3}{AMSa}{36}
      \NewMathSymbol{\geqslant}{3}{AMSa}{3E}

      \let\leq=\leqslant 
      \let\geq=\geqslant 
    \fi
  \fi
\fi % End of OFSS

\ifnfssone
  \newmathalphabet{\mathit}
  \addtoversion{normal}{\mathit}{cmr}{m}{it}
  \addtoversion{bold}{\mathit}{cmr}{bx}{it}
  \newmathalphabet{\mathbfit} % math mode version of \textbfit{..}
  \addtoversion{normal}{\mathbfit}{cmr}{bx}{it}
  \addtoversion{bold}{\mathbfit}{cmr}{bx}{it}
  \newmathalphabet{\mathbfss} % math mode version of \textbfss{..}
  \addtoversion{normal}{\mathbfss}{cmss}{bx}{n}
  \addtoversion{bold}{\mathbfss}{cmss}{bx}{n}
  \ifAMStwofonts
    \ifCUPmtlplainloaded \else
      \UseAMStwoboldmath
      \makeatletter
      \new@mathgroup\upmath@group
      \define@mathgroup\mv@normal\upmath@group{eur}{m}{n}
      \define@mathgroup\mv@bold\upmath@group{eur}{b}{n}
      \edef\UPM{\hexnumber\upmath@group}
      \new@mathgroup\amsa@group
      \define@mathgroup\mv@normal\amsa@group{msa}{m}{n}
      \define@mathgroup\mv@bold\amsa@group{msa}{m}{n}
      \edef\AMSa{\hexnumber\amsa@group}
      \makeatother
      \mathchardef\upi="0\UPM19
      \mathchardef\umu="0\UPM16
      \mathchardef\upartial="0\UPM40
      \mathchardef\leqslant="3\AMSa36
      \mathchardef\geqslant="3\AMSa3E

      \let\leq=\leqslant 
      \let\geq=\geqslant 
    \fi
  \fi
\fi % End of NFSS release 1

\ifnfsstwo
  \DeclareMathAlphabet{\mathbfit}{OT1}{cmr}{bx}{it}
  \SetMathAlphabet\mathbfit{bold}{OT1}{cmr}{bx}{it}
  \DeclareMathAlphabet{\mathbfss}{OT1}{cmss}{bx}{n}
  \SetMathAlphabet\mathbfss{bold}{OT1}{cmss}{bx}{n}
  \ifAMStwofonts
    \ifCUPmtlplainloaded \else
      \DeclareSymbolFont{UPM}{U}{eur}{m}{n}
      \SetSymbolFont{UPM}{bold}{U}{eur}{b}{n}
      \DeclareSymbolFont{AMSa}{U}{msa}{m}{n}
      \DeclareMathSymbol{\upi}{0}{UPM}{"19}
      \DeclareMathSymbol{\umu}{0}{UPM}{"16}
      \DeclareMathSymbol{\upartial}{0}{UPM}{"40}
      \DeclareMathSymbol{\leqslant}{3}{AMSa}{"36}
      \DeclareMathSymbol{\geqslant}{3}{AMSa}{"3E}

      \let\leq=\leqslant 
      \let\geq=\geqslant 
    \fi
  \fi
\fi % End of NFSS release 2

\ifCUPmtlplainloaded \else
  \ifAMStwofonts \else % If no AMS fonts
    \def\upi{\pi}
    \def\umu{\mu}
    \def\upartial{\partial}
  \fi
\fi

\title{Two and Three Dimensional Numerical Simulations of Accretion Discs in a Close Binary System}
\author[Makoto Makita, Kenji Miyawaki and Takuya Matsuda]
       {Makoto Makita\thanks{E-mail address:makita@jet.planet.kobe-u.ac.jp}, Kenji Miyawaki and Takuya Matsuda\thanks{E-mail address:matsuda@jet.planet.kobe-u.ac.jp} \\
        Department of  Earth and Planetary Sciences, Kobe University, Kobe 657-8501, Japan}
\date{}

\pagerange{\pageref{firstpage}--\pageref{lastpage}}
\pubyear{}

\begin{document}

\maketitle

\label{firstpage}

\begin{abstract}
We perform 2D and 3D numerical simulations of an accretion disc in a close binary system using the Simplified Flux vector Splitting (SFS) finite volume method.
In our calculations, gas is assumed to be the ideal one, and we calculate the cases with $\gamma=$1.01, 1.05, 1.1 and 1.2.
The mass ratio of the mass-losing star to the mass-accreting star is one.
Our results show that spiral shocks are formed on the accretion disc in all cases.
In 2D calculations we find that the smaller $\gamma$ is, the more tightly the spiral winds.
We observe this trend in 3D calculations as well in somewhat weaker sense.

Recently, Steeghs, Harlaftis \& Horne (1997) found the first convincing evidence for spiral structure in the accretion disc of the eclipsing dwarf nova binary IP Pegasi using the technique known as Doppler tomography.
We may claim that spiral structure which we have discovered in numerical simulations before are now found observationally. 
\end{abstract}


\section{Introduction}
 One of the main longstanding problems of accretion discs in a close binary system is a mechanism of angular momentum transport. 
Among proposed mechanisms, $\alpha$-model (Shakura \& Sunyaev 1973) is thought to be the standard one.
In this model, viscosity originating from turbulence, magnetism or whatever, is supposed to transport angular momentum.
Nevertheless the origin of the viscosity has been poorly understood so far.

One of other models is the spiral shock model which was first proposed by one of the present authors (Sawada, Matsuda \& Hachisu 1986a, b; Sawada et al. 1987).
In this model the tidal force due to a companion star excites stationary spiral shock waves in the accretion disc. 
The gas in the disc loses its angular momentum when it passes through the spiral shocks.

Sawada et al. (1986a, b, 1987) carried out 2D inviscid calculations of a Roche lobe overflow in a semi-detached binary system with mass ratio of unity.
Their computational region covered both a mass-losing star filling its critical Roche lobe and a mass-accreting compact star.
They used the second-order-accurate Osher scheme with a generalized curvilinear coordinate. 
They found spiral shocks in an accretion disc for the first time.

Since then, a number of two-dimensional simulations of accretion discs have been performed mainly with finite difference/volume method, and they have confirmed that spiral shocks appear in inviscid discs (Spruit et al. 1987; Spruit 1989; Matsuda et al. 1990; Rozyczka \& Spruit 1989; Savonije, Papaloizou \& Lin 1994).
Spruit (1987) obtained self-similar solutions of spiral shocks by a semi-analytic manner (see also Chakrabarti 1990).
He pointed out that the smaller specific heat ratio $\gamma$ is, the more tightly the spiral arms winds. 
Godon (1997) studied tidal effects in accretion discs using a two-dimensional time-dependent hybrid Fourier-Chebyshev spectral method, under the assumptions of a polytropic equation of state and a standard alpha viscosity prescription.
He found a critical value of the viscosity ($\alpha \approx 0.01$), below which the two-armed spiral shock propagates all the way to the inner boundary in cold discs (Mach number ${\cal M} \approx 40$) for the first time.
 
On the other hand, in the case of 3D, this spiral shock model has been criticized on the grounds that the tidally-induced perturbations may find it hard to propagate all the way inwards, due to refraction and diversion into the vertical (stratified) direction (Lin, Papaloizou \& Savonije 1990a, b; Lubow \& Pringle 1993).

Three-dimensional simulations have been carried out mainly using particle methods (Molteni, Belvedere \& Lanzafame 1991; Hirose, Osaki \& Mineshige 1991; Nagasawa, Matsuda \& Kuwahara 1991; Lanzafame, Belvedere \& Molteni 1992, 1993), but their results have not shown the spiral shocks. 
Molteni et al. (1991) and Lanzafame et al. (1992, 1993) stressed that the accretion disc itself was not formed for $\gamma\geq 1.1-1.2$.

Recently, Yukawa, Boffin \& Matsuda (1997) made 3D numerical simulations using Smoothed Particle Hydrodynamics method (SPH) for three values of $\gamma$, i.e. $\gamma$=1.01, 1.1 and 1.2, with mass ratio of unity.
They used $(5-6)\times10^4$ particles, which were much more than those used by previous authors.
They used a variable smoothing length technique.
They demonstrated that the accretion disc as well as spiral shocks existed in the case of $\gamma=1.2$, contrary to previous claims, although they could not obtain clear spiral shocks in the cases of $\gamma$=1.1 and 1.01.

Lanzafame \& Belvedere (1997, 1998) performed 3D SPH calculations of an accretion disc in a close binary system via wind accretion mechanism.
Their simulations were made for a Her X-1--HZ Her like system with a primary mass $M_1 = 1.3~M_{\odot}$ and a secondary mass $M_2 = 2.2~M_{\odot}$ (1997) and  for a Cen X-3 like system with $M_1 = 1.4~M_{\odot}$ and $M_2 = 19.1~M_{\odot}$ (1998).
They considered quasi-polytropic gas ($p \propto \rho^{\gamma}$) with $\gamma=1.01$.
In their calculation, they calculated the region including  a portion of the secondary Roche lobe surface.
They used 55,747 (1997) and 113,676 (1998) SPH particles.
They found that the azimuthal distribution of the radial Mach number showed spiral shocks. 
In a Her X-1--HZ Her case (1997), shocks appeared only in the outer regions.
On the other hand in a Cen X-3 case (1998), three spiral shocks persisted from the outer edge to the inner regions.   

In the case of finite difference/volume method, there had been only one three-dimensional calculation (Sawada \& Matsuda 1992) for a while. 
They used a Roe upwind TVD scheme to calculate the case of $\gamma=1.2$ with mass ratio of unity.
They found the existence of spiral shocks on the accretion disc, but their calculation was done up to only half a revolution period.
It might be argued that their computational time is not long enough to obtain a convincing result.

In a series of papers Bisikalo et al. have done 3D numerical studies of the flow structure in semi-detached binary systems (1997a, b, 1998a, b, c).
They used a TVD Roe scheme modified by monotonic flux limiters in the Osher's form.   
Their computational region was a parallelepipedon $[-a .. 2a] \times [-a .. a] \times [ 0 .. a]$, where $a$ is the orbital separation.
Non-uniform difference grids  (finer near the accretor) containing $78 \times 60 \times 35 $ grid points for the system X1822-371 and $84 \times 65 \times 33$ grid points for the system Z Cha were used (1998b).
Their calculations had been stopped at 12-20 revolution periods.
In their calculations, gas stream from L1 did not cause the shock perturbation of the disc boundary.
It meant the absence of a hot spot.
Their results also showed that the gas of 'circumbinary envelope' interacted with the stream and caused the formation of an extended shock wave, located on the stream edge. 
However, their shock waves did not form clear spiral structure.
They also claimed that an accretion disc was not formed for the case of $\gamma=1.2$ (1998c) and it was very elongated for $\gamma=1.01$. 
As is stated above, not only the existence of spiral shocks but that of the disc itself (for higher $\gamma$) in 3D has not been established convincingly.

From observational view points, recently, Steeghs et al. (1997) found the first convincing evidence for spiral structure in the accretion disc of the eclipsing dwarf nova binary IP Peg at the outburst phase using the technique known as Doppler tomography.

Neustroev \& Borisov (1998) studied the dwarf nova U Geminorum and claimed that they obtained the convincing evidence for the existence of spiral shocks in the accretion disc of a cataclysmic variable in quiescence.

The purpose of the present paper is to perform 2D and 3D finite volume numerical calculations with higher resolution up to enough time to confirm the existence of an accretion disc and spiral shocks on it. We also stress that the spirals are induced by the tidal force due to the companion.

\section{Method of calculations}
\subsection{Basic assumptions}
We consider a binary system composed of a mass-accreting primary star with mass $M_1$ and a mass-losing secondary star with $M_2$.
The mass ratio is defined as $q=M_2/M_1$.
In the present calculations the mass ratio is fixed to be one.

We solve only the region around the mass-accreting star. 
The gas is assumed to be supplied at the L1.
In 2D calculations, we assume that the flow is confined within a narrow gap in the orbital plane with a constant thickness, so that these calculations become two-dimensional.
We assume an ideal gas with constant specific heat ratio $\gamma$. 
Although we do not include heating and cooling effects explicitly, we choose the smaller $\gamma$ so that these effects are implicitly taken into account.
In the present paper, we examine the cases with $\gamma$=1.2, 1.1, 1.05 and 1.01.
These calculations contain no viscosity except numerical dissipation.

In our 3D calculations, the assumptions of calculations are the same as in 2D calculations except that gas is not restricted within a gap of constant thickness.
Symmetry about the orbital plane is assumed.

\subsection{Basic equations}
The basic equations we are going to solve are the two-dimensional as well as the three-dimensional Euler equations describing the motion of an inviscid gas.

We choose the orbital separation, $a$, as the length scale, and $a~\omega$ as the velocity scale, where $\omega$ is the angular velocity of the system.
Therefore, the revolution period is normalized to $2\pi$.
Density is normalized by the density of the gas at L1 point. 
The primary star is centred at the origin and the secondary star at (-1, 0, 0).
We set a numerical grid centred at the origin and just touching the L1 point.
Cartesian coordinates we use have $200\times200$ grid points in 2D and $200\times200\times50$ grid points in 3D, respectively.

Our 3D basic equations in the rotating frame of reference can be described in dimensionless form as 
\begin{equation}
\frac{\partial \tilde{Q}}{\partial t}+\frac{\partial\tilde{E}}{\partial x}+\frac{\partial\tilde{F}}{\partial y}+\frac{\partial\tilde{G}}{\partial z}+\tilde{H}=0, 
\label{eqn:euler3}
\end{equation}
\[
\tilde{Q} =\left(
\begin{array}{c}
\rho \\
\rho u \\
\rho v \\
\rho w \\
e
\end{array}\right),\\
\quad
\tilde{E} = \left(
\begin{array}{c}
\rho u \\
\rho u^2+p \\
\rho u v \\
\rho u w \\
(e+p) u
\end{array}\right),\\
\quad
\tilde{F} =\left(
\begin{array}{c}
\rho v \\
\rho u v \\
\rho v^2+p \\
\rho v w \\
(e+p) v
\end{array}
\right),\\
\]
\begin{equation}
\tilde{G} =\left(
\begin{array}{c}
\rho w \\
\rho u w \\
\rho v w \\
\rho w^2+p \\
(e+p) w
\end{array}
\right), \\
\quad
\tilde{H} = \left(
\begin{array}{c}
0 \\
\rho\kappa_x \\
\rho\kappa_y \\
\rho\kappa_z \\
\rho(u~\kappa_x+v~\kappa_y+w~\kappa_z)\\
\end{array}\right),
\end{equation}
where $\rho, p, e,$ $u$, $v$, $w$, $\kappa_x$, $\kappa_y$, and $\kappa_z$ are the density, the pressure, the total energy per unit volume, the $x$, $y$ and $z$ components of velocity and the $x$, $y$, and $z$ components of force, respectively.
The $x$, $y$ and $z$ components of  force are
\begin{eqnarray}
\kappa_x&=& -2 v -x_{\rm G}+\frac{1/(1+q)}{r^3_1}x_1+\frac{q/(1+q)}{r^3_2}x_2, 
\label{eqn:force1}
\\
\kappa_y&=&2 u-y+\frac{1/(1+q)}{r_1^3}y_1+\frac{q/(1+q)}{r^3_2}y_2,
\label{eqn:force2}
\\
\kappa_z&=& \frac{1/(1+q)}{r^3_1}z_1+\frac{q/(1+q)}{r^3_2}z_2. 
\label{eqn:force3}
\end{eqnarray} 
Here, $r_i (i=1, 2)$ is the distance from the point considered to the centre of each star, and $x_i$, $y_i$ and $z_i$ are each star's Cartesian components.
$x_{\rm G}$ represents the $x$ component of the distance from the centre of mass.

\begin{figure}
\begin{center}
\leavevmode
\epsfxsize=8cm
\epsfbox{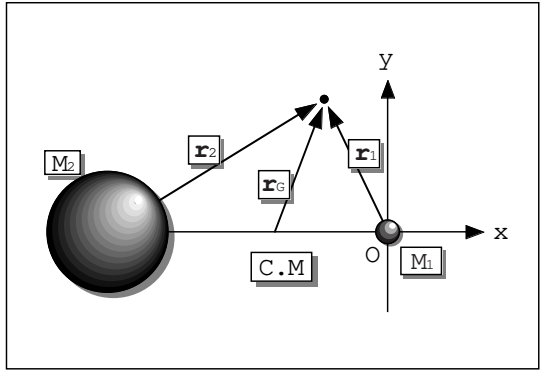}
\caption{Geometry of the system.}
\label{fig:geom}
\end{center}
\end{figure}

In our calculations, the mass-accreting star is centred at the origin, and the mass-losing star at (-1, 0, 0).
Hence 
\begin{eqnarray}
x_1 &=& x,  \\
x_2 &=& x+1, \\
x_{\rm G} &=& x+\frac{q}{1+q}.
\end{eqnarray}
This geometry is shown in Fig. \ref{fig:geom}.

In equations (\ref{eqn:force1}) and (\ref{eqn:force2}), the first term represents the Coriolis force, the second one the centrifugal force, respectively.
The last two terms in equations (\ref{eqn:force1}) and (\ref{eqn:force2}) as well as ones in equation (\ref{eqn:force3}) represent the gravity force.

The equation of state we use is
\begin{equation}
p=(\gamma-1)\left[e-\frac{\rho}{2}(u^2+v^2+w^2)\right].
\end{equation}

In the normalization procedure, we use an auxiliary equation: 
\begin{equation}
a^3 \omega^2 = G(M_1+M_2),
\end{equation} 
to eliminate the gravitational constant $G$ from the equations.

\subsection{Numerical method}
We use Simplified Flux vector Splitting (SFS) scheme (Jyounouti et al. 1993; Shima \& Jyounouti 1994, 1997) (see Appendix). 
In order to obtain high spatial resolution we use a MUSCL (Monotone Upwind Scheme for Conservation Law) type approach.
The numerical accuracy of the method is the second-order both in space and in time.

\subsection{Initial and boundary conditions}
As an initial condition we suppose the whole computational region, which is $-0.5 \leq x, y \leq 0.5$ in 2D and furthermore $0 \leq z \leq 0.25$ in 3D, is filled with tenuous gas with higher temperature at the initial instance.
We assume $\rho_0=10^{-5},~p_0=10^{-4}/ \gamma$ and $u_0=v_0=w_0=0$ in the whole numerical region.

This initial condition is also maintained outside of the outer boundary except for L1 point and inside of the inner boundary during calculations. This choice forms the outer and the inner boundary conditions except the inlet at L1 point.

At L1 point a small rectangular inlet hole is placed. The values at the inlet  are $\rho_{in} = 1.0,~p_{in} = 10^{-2}/\gamma,~u_{in} = 0.01$ and $v_{in} = w_{in} = 0.0$ at all the time.
This choice of the  density and the pressure means that the sound speed of the gas to be injected into the computational region is $10^{-1}$.
Therefore, we may say that the temperature of the gas is very high.
This is necessary to ensure a sufficient amount of gas flux from the inlet.
The mass flux from L1 point is computed by solving a Riemann problem, so the amount of inflow gas is not the same between 2D and 3D simulations. 

The mass-accreting star is represented by a hole of $3 \times 3$ cells in 2D cases and $ 3 \times 3 \times 2$ cells in 3D cases situated at the origin.
We follow the time evolution until $t = 20\pi$ (that is 10 revolution periods) in 2D calculations and $t = 10\pi$ in 3D cases.

\section{Results}
\subsection{2D calculations}
\begin{figure*}
\begin{center}
\leavevmode
\epsfxsize=12cm
\epsfbox{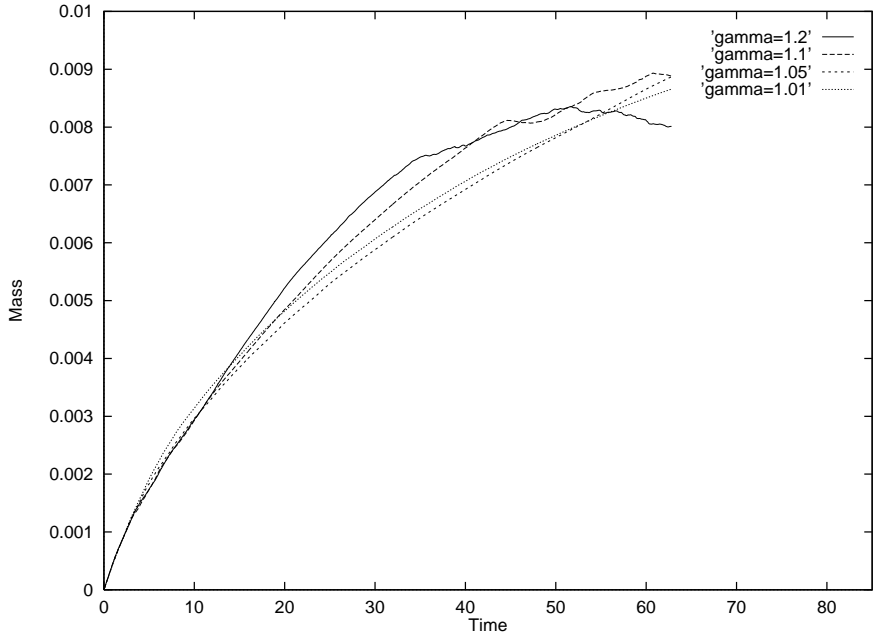}
\end{center}
\caption[ma2d]{Evolution of the disc mass as a function of time.}
\label{fig:mass2d}
\end{figure*}

The time evolution of the disc mass is shown in Fig. \ref{fig:mass2d}.
In the case of $\gamma=1.2$, the mass is saturated at about $t=50$.
The disc mass is still increasing except for the case of $\gamma=1.2$. 
We have not yet reached a steady state in the other cases.
In order to investigate a time variation of flow, we make video movies.
From the movies we confirm that a gross feature of the flow in our calculations does not change except minor oscillations.
We are confident that we have almost obtained the final structure of the disc.
 
\begin{figure*}
\begin{center}
\begin{tabular*}{1.0\columnwidth}{p{0.5\columnwidth}p{0.5\columnwidth}}
\leavevmode
\epsfxsize=72mm
\epsfbox{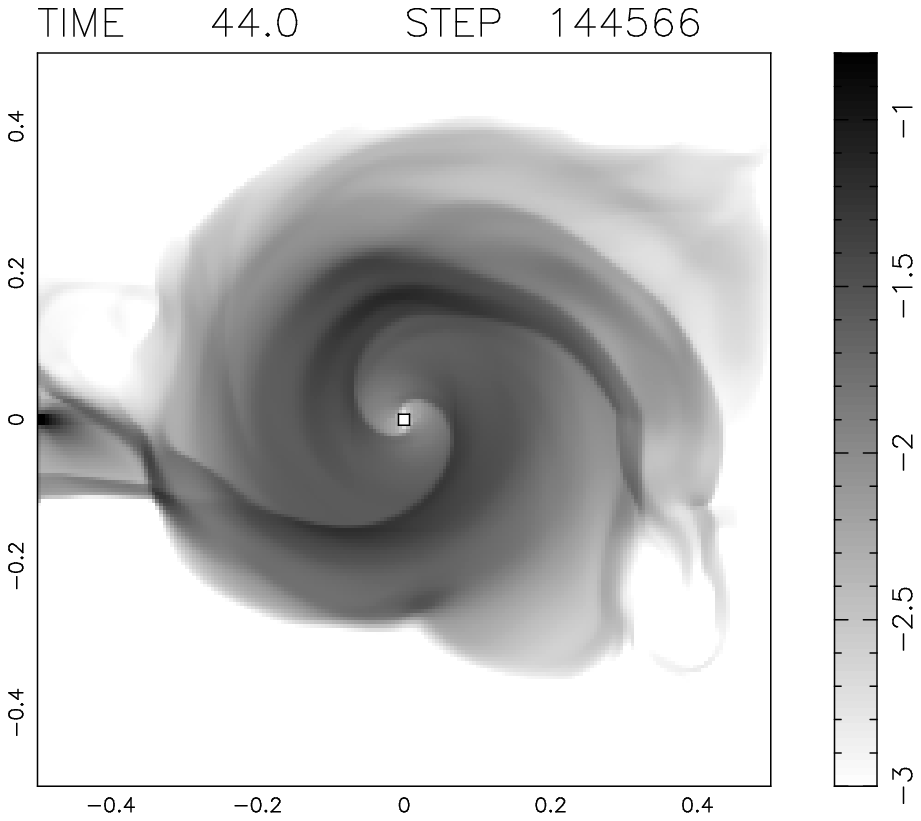}&
\epsfxsize=72mm
\epsfbox{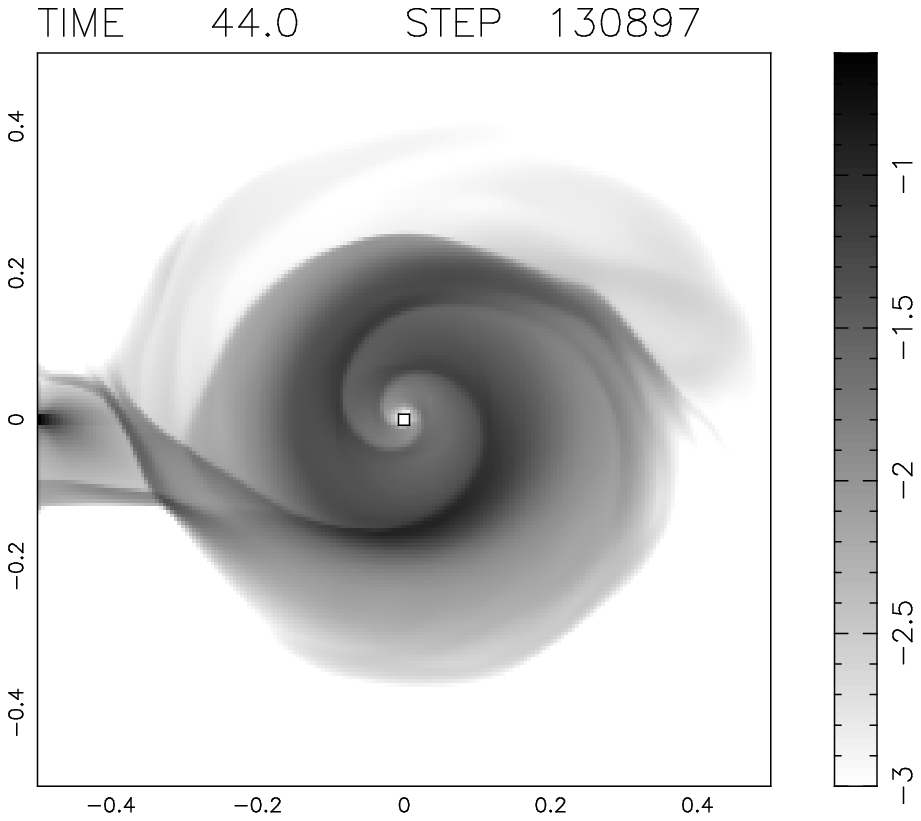}\\
\vspace{-5mm}\hspace{25mm}{\Large$\gamma=1.2$}&
\vspace{-5mm}\hspace{25mm}{\Large$\gamma=1.1$}\\
\epsfxsize=72mm
\epsfbox{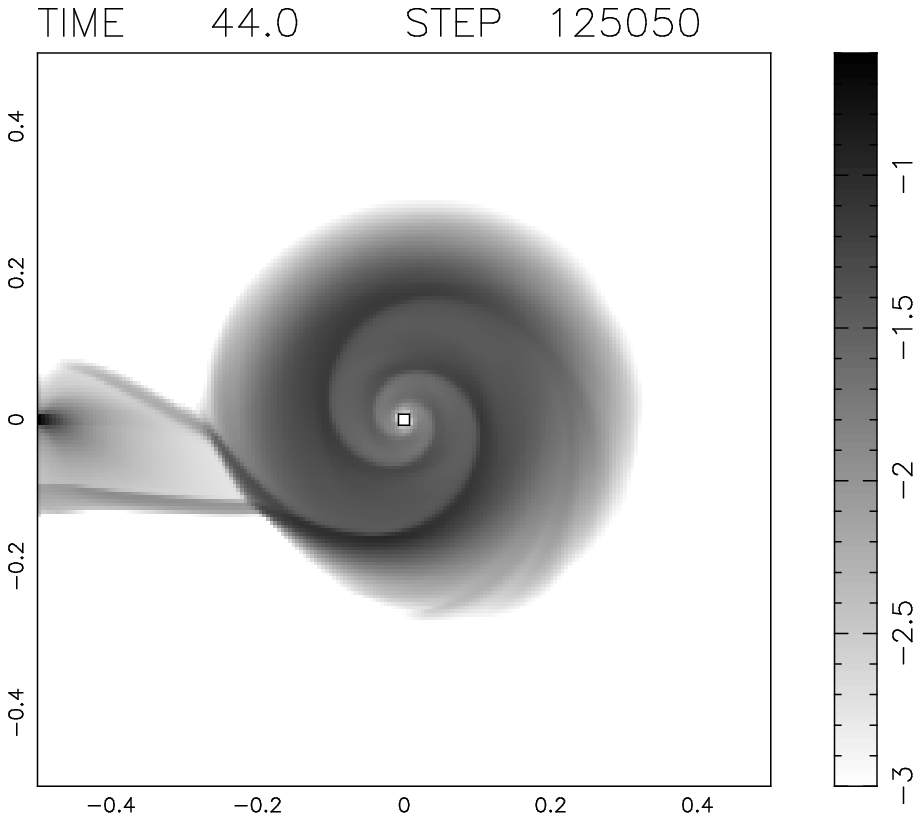}&
\epsfxsize=72mm
\epsfbox{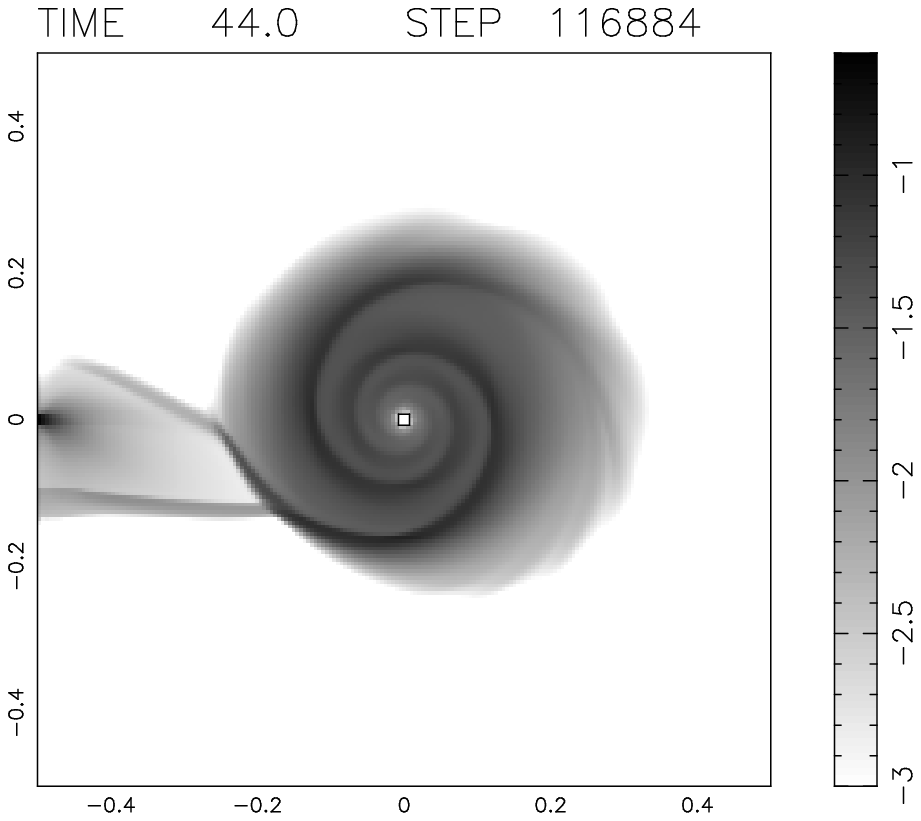}\\
\vspace{-5mm}\hspace{2.5cm}{\Large$\gamma=1.05$}&  
\vspace{-5mm}\hspace{2.5cm}{\Large$\gamma=1.01$}\\
\end{tabular*}
\end{center}
\caption[d2d]{Gray scale of density distribution with logarithmic scale at 7 revolution periods. Bar in the right side shows the scale range.}
\label{fig:dense2d}
\end{figure*}

We show the gray scale of density distribution for all cases at $t=44$, which is 7 revolution periods, in Fig. \ref{fig:dense2d}. 
We find two clear spiral shocks in these figures as the same as the results of the previous authors.
The inflow from L1 point expands rapidly because of rather high temperature of the gas assumed.
This flow is a kind of an under-expanded jet; we observe two intercepting shocks are formed.
These shocks are artifact due to our boundary condition at L1 point.
In order to remove these shocks, we have to solve the whole region containing the mass-losing star as was done by Sawada et al. (1986a, b, 1987). 
However, we believe that the structure of the accretion disc is not affected by the present choice of the boundary condition at L1 point.
It can be proved by comparing the present results with our previous ones (Sawada et al. 1986a, 1987).

From the point of view of dependence on $\gamma$, we find that the spiral arms wind more tightly in smaller $\gamma$ cases than larger ones.
Lower $\gamma$ means a cooler disc with larger Mach number of the flow.
Our results coincide with those of Savonije et al. (1994) and Godon (1997), and the analytical work of Spruit (1987).
Spruit also argued that eventually spiral waves become a ring for $\gamma=1$.
Our results show that spiral shocks still exist even for $\gamma=1.01$.
However, our disc has a constant thickness, while Spruit considered a tapered disc.
Therefore, direct comparison may not be appropriate. 

We also observe that the flow for higher $\gamma$ are more unsteady in time, of which fact may be observed in the case of $\gamma=1.2$ in Fig. \ref{fig:dense2d} and also in our video movies.

\subsection{3D calculation}
\begin{figure*}
\begin{center}
\begin{tabular*}{1.0\columnwidth}{p{0.5\columnwidth}p{0.5\columnwidth}}
\leavevmode
\epsfxsize=72mm
\epsfbox{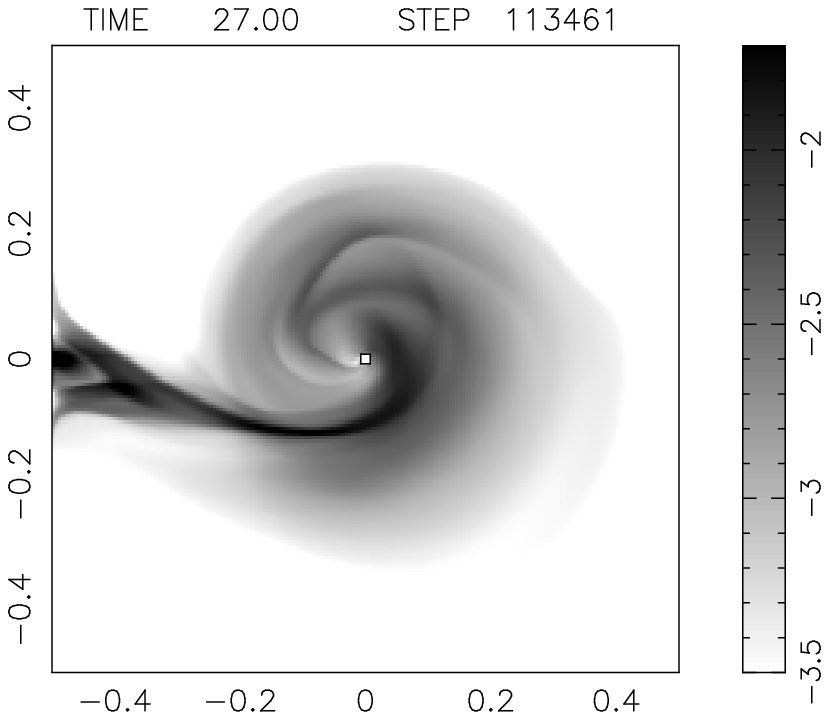}&
\epsfxsize=72mm
\epsfbox{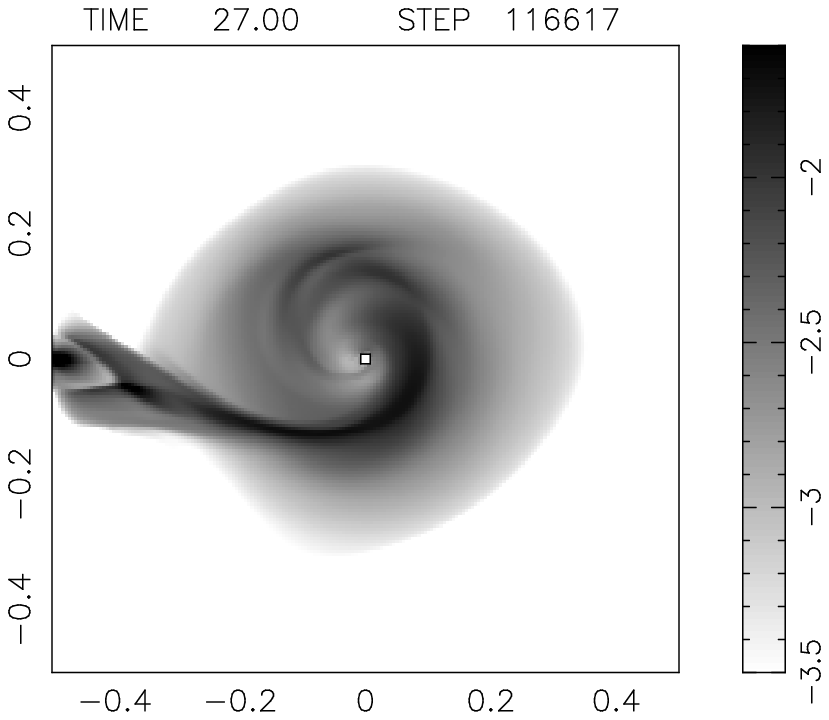}\\
\vspace{-5mm}\hspace{25mm}{\Large $\gamma=1.2$}&
\vspace{-5mm}\hspace{25mm}{\Large$\gamma=1.1$}\\
\epsfxsize=72mm
\epsfbox{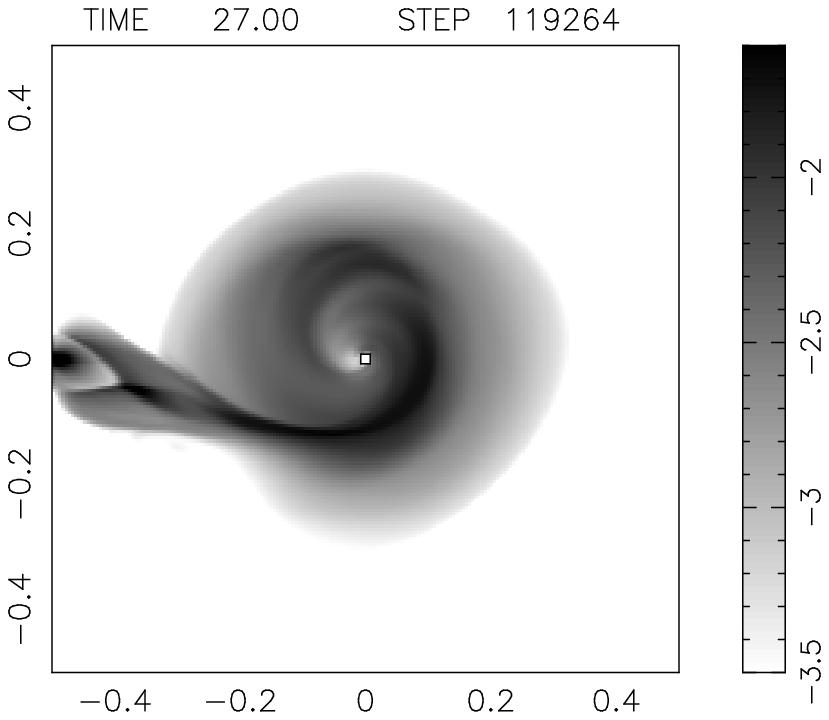}&
\epsfxsize=72mm
\epsfbox{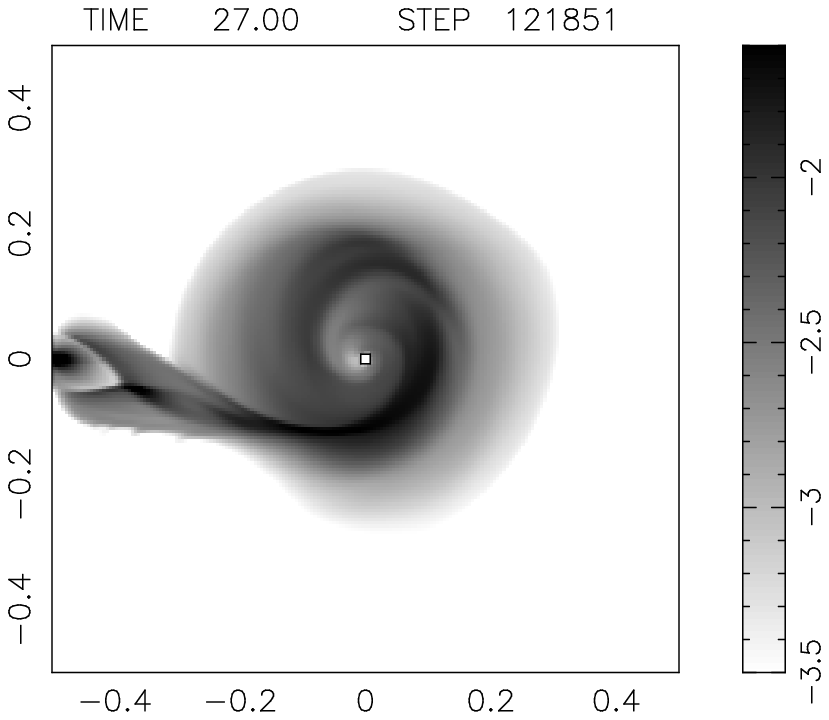}\\
\vspace{-5mm}\hspace{2.5cm}{\Large$\gamma=1.05$}& 
\vspace{-5mm}\hspace{2.5cm}{\Large$\gamma=1.01$}\\
\end{tabular*}
\end{center}
\caption[dxy]{Gray scale of density distribution with logarithmic scale in the orbital plane at $t=27$.}
\label{fig:densex-y}
\end{figure*}

We present density profiles in the orbital plane in all four cases in Fig. \ref{fig:densex-y}.
These figures show the existence of spiral shocks in not only the case of $\gamma=1.2$ but also in the other three cases. 
We note the remarkable similarity between the present case of $\gamma=1.2$ and those by Sawada et al. (1992) and  by Yukawa et al. (1997). 
However, Yukawa et al. (1997) could not obtain spiral structures for $\gamma=1.1$ and $1.01$.
We may attribute this to the lack of number of particles in their calculations.
As to the accuracy of the computational results, we would point out that our number of grid points is $4 \times 10^6$ for the whole computational domain compared with $(5-6) \times 10^4$ SPH particles in Yukawa et al. (1997). 
Although it may not be appropriate to compare these numbers directly, the difference of two orders of magnitude in these numbers may be significant. 
In fact Lanzafame \& Belvedere (1998) observed spiral structures for the case of $\gamma=1.01$ using their SPH code with 113,676 particles. Therefore, we believe that there should exist spiral structures on the accretion discs as long as $\gamma \leq 1.2$.

Comparing our case of $\gamma=1.01$ with our case of $\gamma=1.2$, we see that the former case shows slightly tightly wound spiral shocks near the centre.
We, however, do not find a clear difference in spiral structure in the outer region particularly for the cases of $\gamma \leq 1.1$.
We, therefore, may conclude that  the dependence of spiral structure on the value of $\gamma$ in 3D calculation is somewhat weaker than that in the 2D cases.

We make video movies for all cases and confirm that the flow reaches almost steady state.
It is observed that the higher $\gamma$ cases are more violent than the lower $\gamma$ cases, of which fact is the same as in 2D cases.

We provide the three-dimensional views of the iso-density surface of $\rho=10^{-3}$ in Fig. \ref{fig:isodense}.
Fig. \ref{fig:densex-y} may lead an impression that the stream from L1 point goes into the disc directly without the interaction between the stream and the disc. However, this is certainly not the case. Fig. \ref{fig:isodense} clearly shows that the stream does not penetrate into but collides with the disc.
Therefore we do not confirm the claim by Bisikalo et al. (1998b, c) on the non-existence of a so-called hot spot.   
\begin{figure}
\begin{center}
\leavevmode
%\epsfxsize=72mm
%\epsfbox{g12-45.eps}
\end{center}
\begin{center}
\leavevmode
%\epsfxsize=72mm
%\epsfbox{g101-45.eps}
\end{center}
\caption[iso]{3D view of iso-density surface at the level of $\rho = -3$ with logarithmic scale at 5 revolution periods. Gray scale of density in the orbital plane is also shown. Bar in the left side shows the scale range.}
\label{fig:isodense}
\end{figure}

In Fig. \ref{fig:isodense} we observe the accretion disc, the spiral structures in it and the stream from L1 point. We stress here again that the accretion disc does exist for the case of $\gamma=1.2$ contrary to the claims by Lanzafame et al. (1992) and Bisikalo et al. (1998c). Our result is consistent with our earlier finite volume calculation by Sawada et al. (1992) and the SPH calculation by Yukawa et al. (1997).

The strange shape of the stream from L1 point is due to the fact that the shape of the inlet at the L1 point is rectangular rather than circular, and the stream is an under expanded jet. 
Nevertheless, it is important to stress again that the stream is an independent object from the accretion disc. 

The cause of the discrepancy between our result and Bisikalo et al. (1998b, c) is not well understood. 
However, we should point out that our number of grid points above the orbital plane is $2 \times 10^{6}$, while theirs is at most $1.8 \times 10^{5}$, i.e. ours is more than 10 times more than theirs. 
Moreover the volume of our computational domain is 0.25, while theirs is 6. It means that the volume of their computational cell is almost 240 times bigger than ours on average, which suggests that their calculations are much cruder than ours. We are not sure if this is the sole reason of the discrepancy, though.
 
\begin{figure*}
\begin{center}
\begin{tabular*}{1.0\columnwidth}{p{0.5\columnwidth}p{0.5\columnwidth}}
\leavevmode
\epsfxsize=72mm
\epsfbox{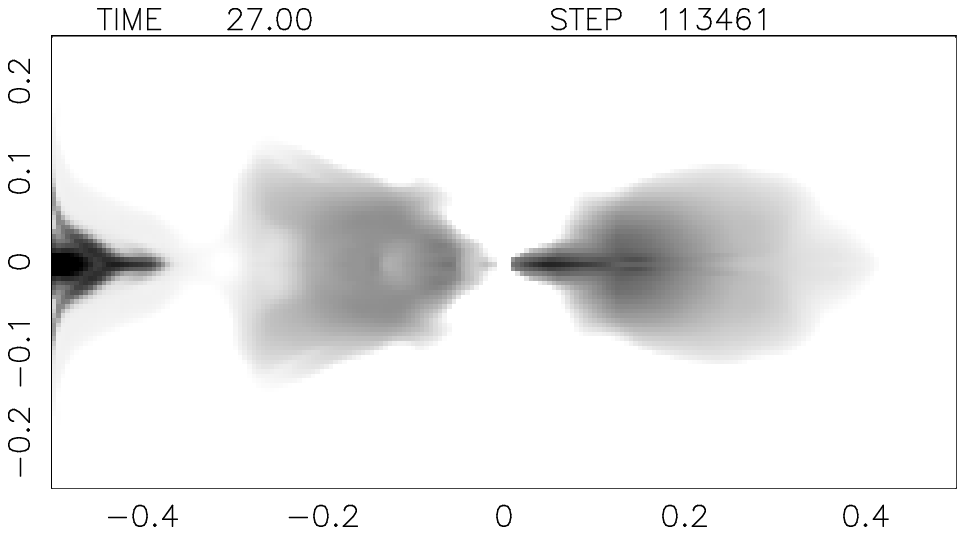}&
\epsfxsize=72mm
\epsfbox{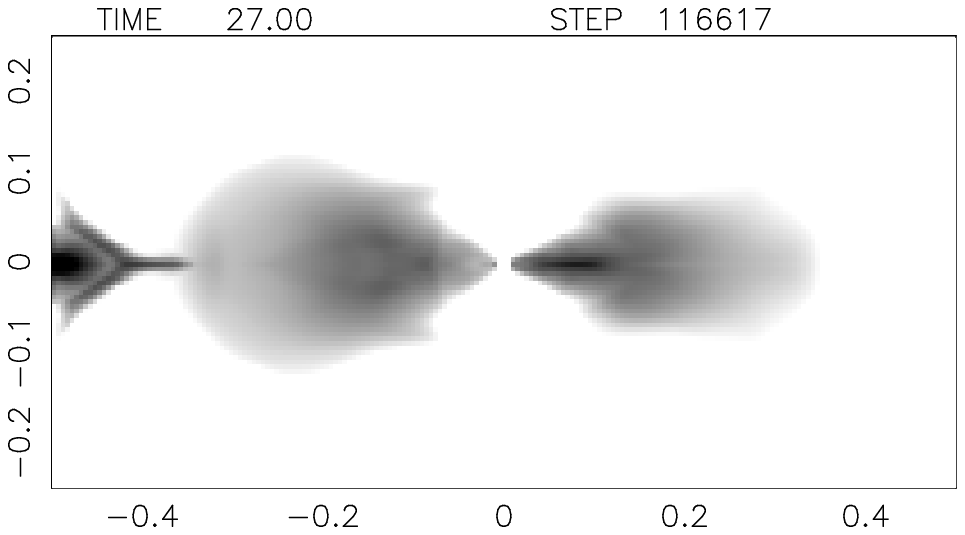}\\
\vspace{-5mm}\hspace{25mm}{\Large $\gamma=1.2$}&
\vspace{-5mm}\hspace{25mm}{\Large$\gamma=1.1$}\\
\epsfxsize=72mm
\epsfbox{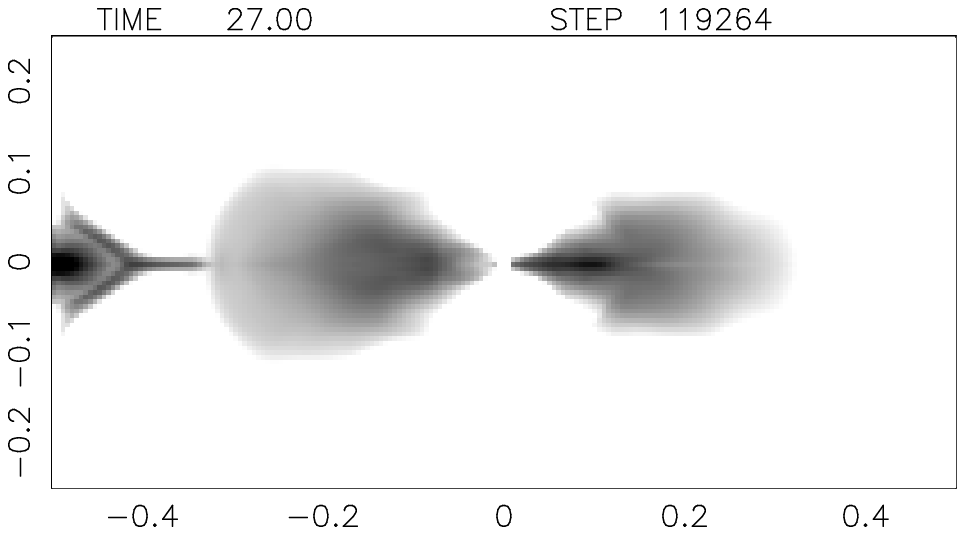}&
\epsfxsize=72mm
\epsfbox{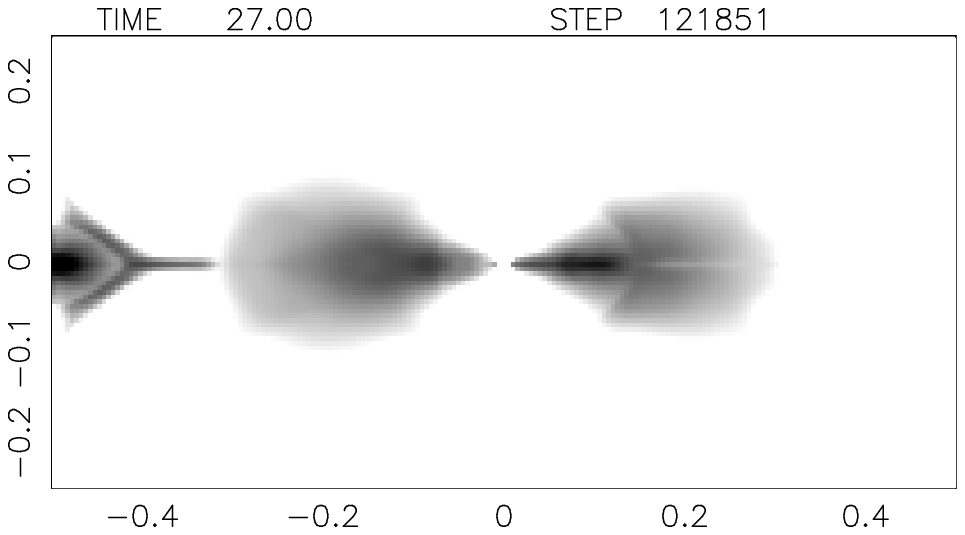}\\
\vspace{-5mm}\hspace{2.5cm}{\Large$\gamma=1.05$} & 
\vspace{-5mm}\hspace{2.5cm}{\Large$\gamma=1.01$} \\
\end{tabular*}
\end{center}
\caption[dxz]{Gray scale of density distribution in the x-z plane at $t=27$.
The scale range is logarithmic one which is the same as Fig. \ref{fig:densex-y}.}
\label{fig:densex-z}
\end{figure*}

Fig. \ref{fig:densex-z} presents the density profiles in the x-z plane.
We can see that disc thickness near the companion star is thicker than the other side.
This feature coincides with the result of Yukawa et al. (1997).
We also find that the smaller $\gamma$ is, the more symmetry the disc structure is.

\begin{figure}
\begin{center}
\leavevmode
\epsfxsize=80mm
\epsfbox{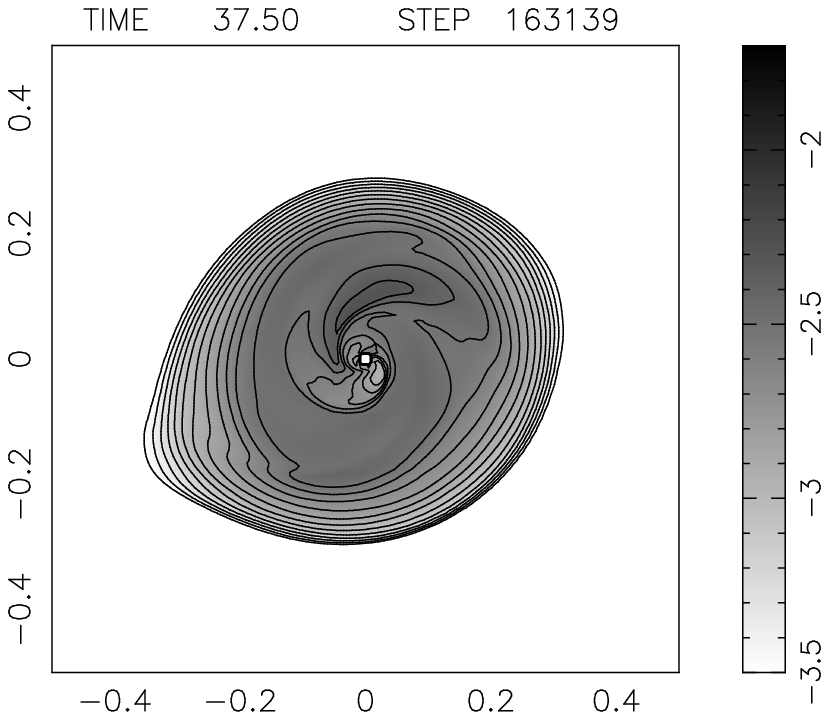}\\
\vspace{5mm}
\hspace{-10mm}
\epsfxsize=80mm
\epsfbox{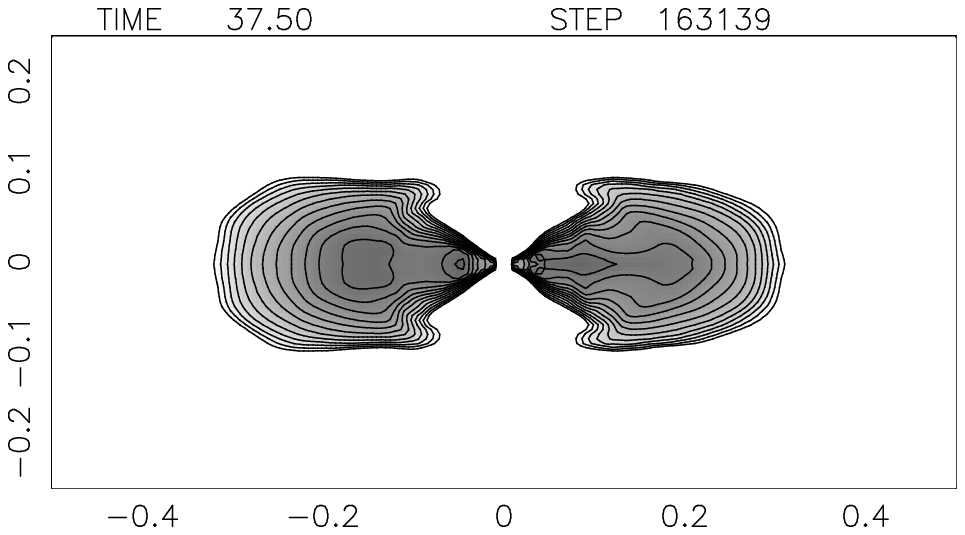} 
\end{center}
\caption[11nixyz]{Gray scale and contours of density profile in the case of $\gamma=1.1$ at $t=37.5$. In this case inflow is stopped at 5 revolution periods. \quad Top: x-y plane, Bottom: x-z plane.}
\label{fig:11ni}
\end{figure}

We make a test calculation in which gas inflow is terminated at $t=10\pi$ to see the effect of the inflow on the spiral shocks.
We show the density profiles in the case of $\gamma=1.1$ at one revolution period after the inflow is terminated, in Fig. \ref{fig:11ni}. 
We observe clear two-armed spiral shocks of bi-symmetry. The spiral shocks are also found even after two revolution period.
We, therefore, conclude that the spiral shocks is formed by tidal force, not by the inflow, of which claim was posed by Bisikalo et al. (1998b, c).

\begin{figure}
\begin{center}
\leavevmode
\epsfxsize=80mm
\epsfbox{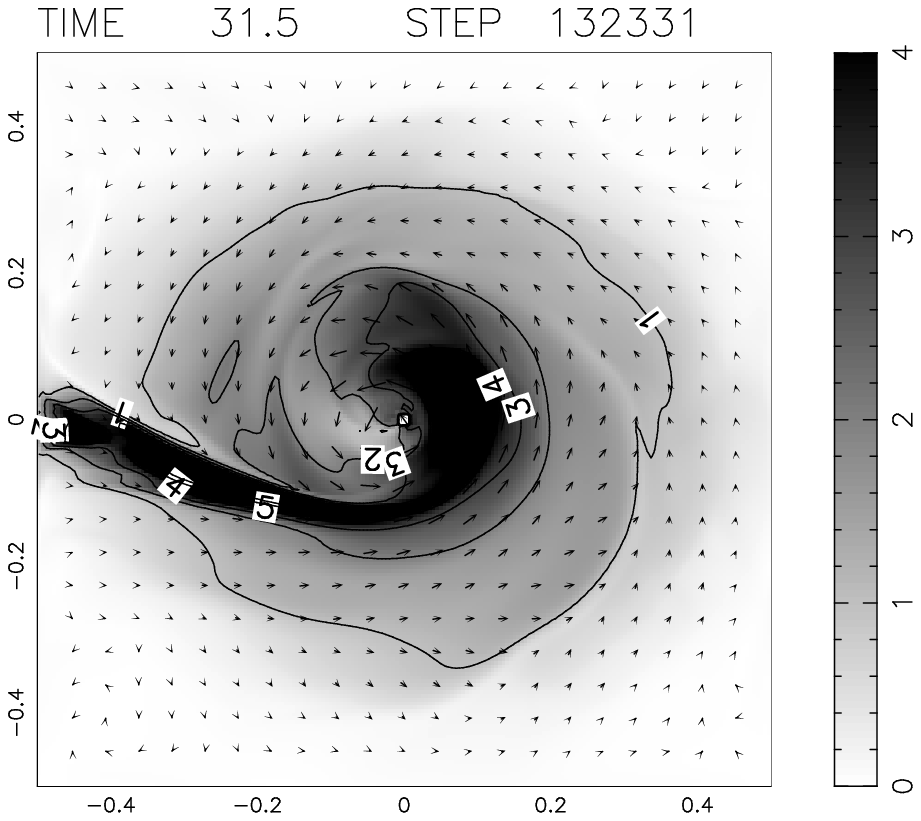}\\
\vspace{5mm}
\hspace{0mm}
\epsfxsize=80mm
\epsfbox{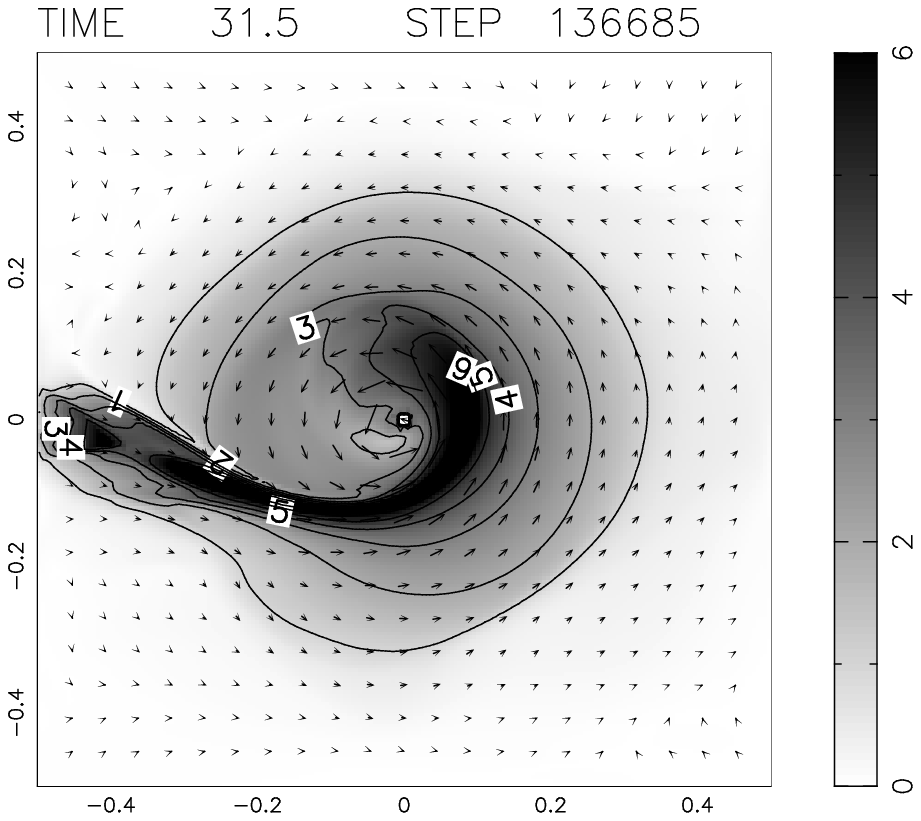}
\end{center}
\caption[machn1]{Mach number contours with gray scale and velocity vectors in the orbital plane at 5 revolution periods. \quad Top: The case of $\gamma=1.2$, Bottom: The case of $\gamma=1.1$.}
\label{fig:12mn11mn}
\end{figure}

\begin{figure}
\begin{center}
\leavevmode
\epsfxsize=80mm
\epsfbox{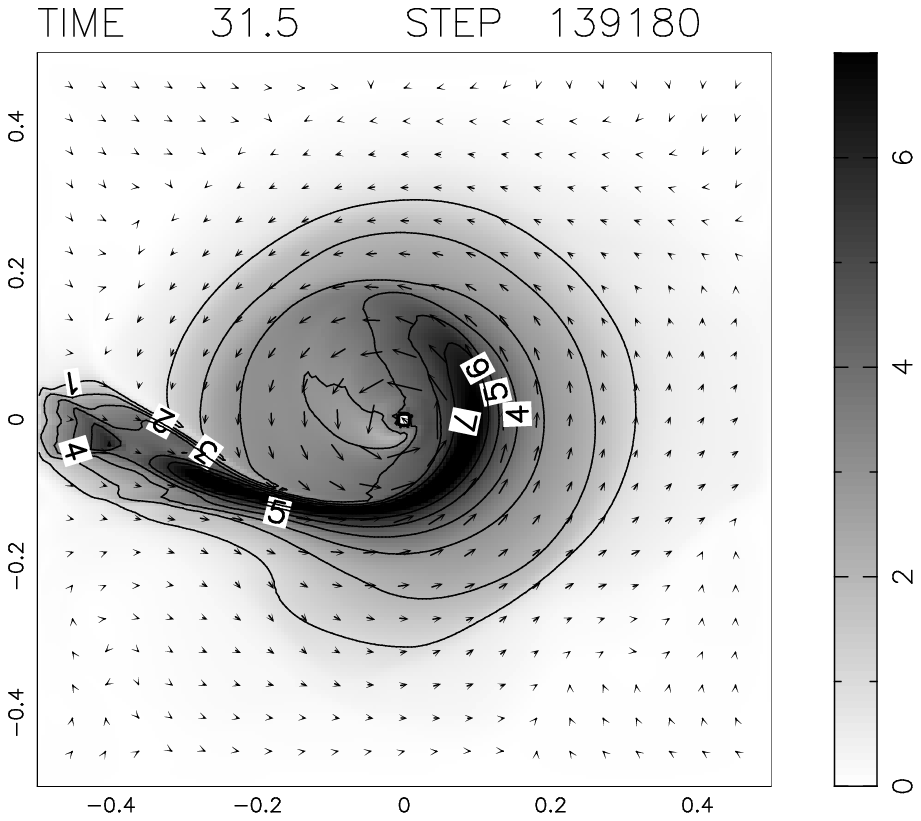}\\
\vspace{5mm}
\hspace{0mm}
\epsfxsize=80mm
\epsfbox{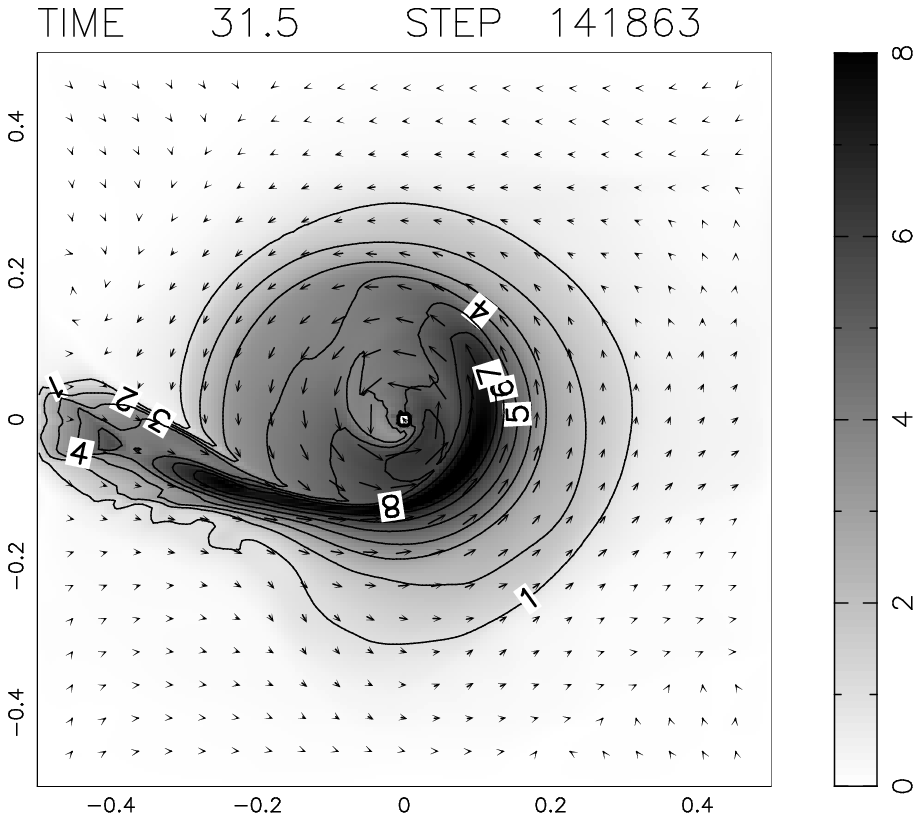}
\end{center}
\caption[machn2]{Mach number contours with gray scale and velocity vectors in the orbital plane at 5 revolution periods. \quad Top: The case of $\gamma=1.05$, Bottom: The case of $\gamma=1.01$.}
\label{fig:105mn101mn}
\end{figure}

Figs. \ref{fig:12mn11mn} and \ref{fig:105mn101mn} show Mach number contours with gray scale as well as the velocity vectors in the orbital plane at 5 revolution periods. 
In these figures, the velocity vectors change the direction abruptly at the positions corresponding to the spirals shown in Fig. \ref{fig:densex-y}. 
We find also that Mach number in accretion discs become higher for smaller $\gamma$.
This feature coincides with the results of previous 2D calculations. 
It is clear that the velocity in the outer region is nearly circular and slower for all cases. This feature in $\gamma=1.2$ case also contradicts very much with Bisikalo et al. (1998c), who found a large mass outflow from the system.

To confirm the existence of spiral shocks and to compare with the observations by Steeghs et al. (1997), we make a Doppler map, which is essentially a hodograph used in hydrodynamics.
Doppler map shows the distribution of a physical variable in velocity coordinates ($V_x, V_y$). 
The relation between the Doppler map and the physical space is as follows:
\begin{enumerate}
\item Doppler map is $90^{\circ}$ rotated anti-clockwise (in our case) from the physical space.
\item The inner region of the accretion disc, which has faster rotational velocity, is mapped to the outer region of the Doppler map, and vice versa.
\item Spiral structure in the physical space is mapped to that in the Doppler map. 
\end{enumerate}
From the property (iii) we may say that, if spiral structure is observed in a Doppler map, it shows the existence of spiral structure in a physical space.
(See Marsh \& Horne (1988), Steeghs et al. (1996) and Steeghs et al. (1997)). 
This is the basis of their claim of the discovery of spiral structure in an accretion disc in IP Peg by Steeghs et al. (1997).

\begin{figure*}
\begin{center}
\begin{tabular*}{1.0\columnwidth}{p{0.5\columnwidth}p{0.5\columnwidth}}
\leavevmode
\epsfxsize=72mm
\epsfbox{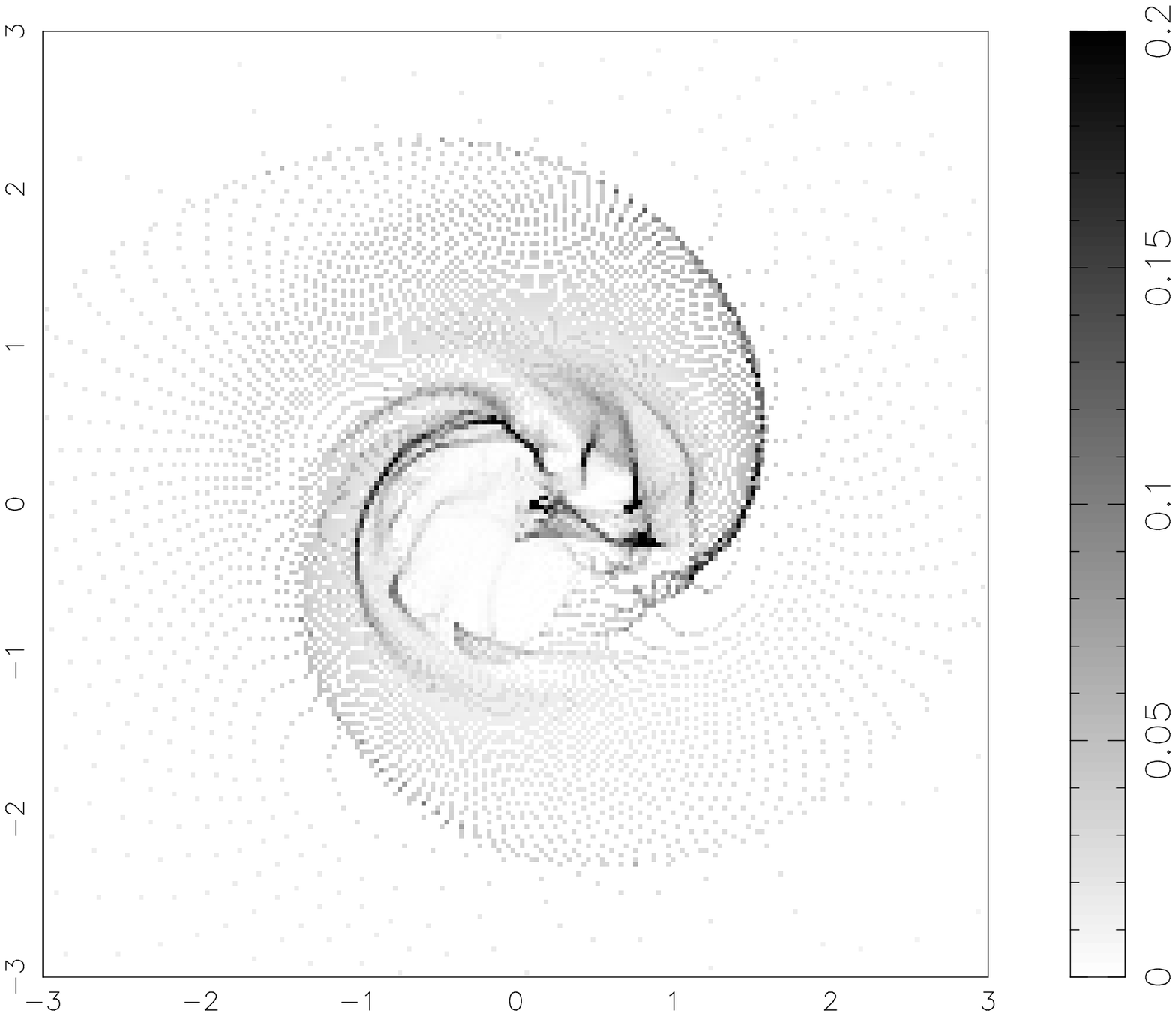}&
\epsfxsize=72mm
\epsfbox{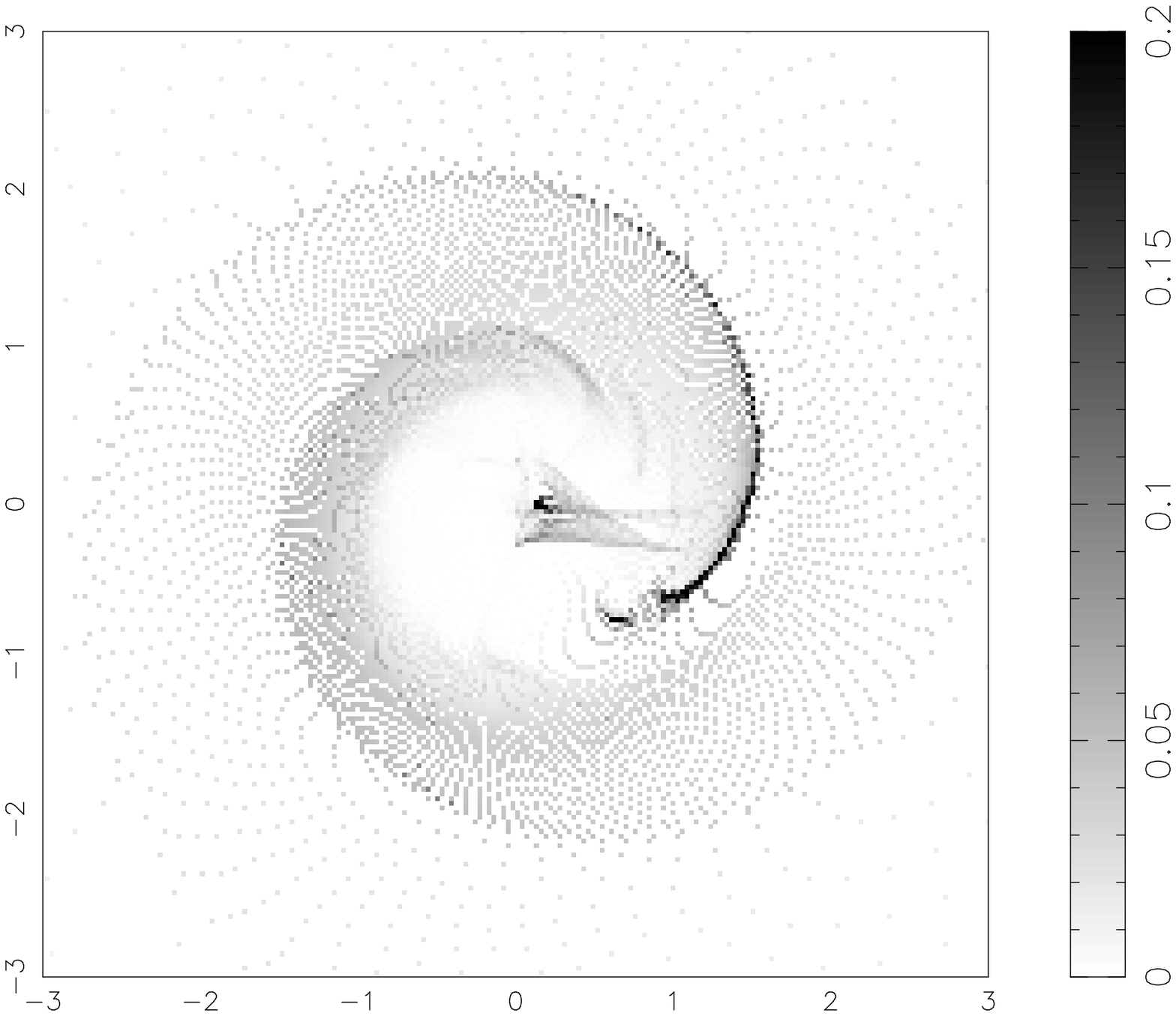}\\
\vspace{-5mm}\hspace{25mm}{\Large $\gamma=1.2$}&
\vspace{-5mm}\hspace{25mm}{\Large $\gamma=1.1$}\\
\epsfxsize=72mm
\epsfbox{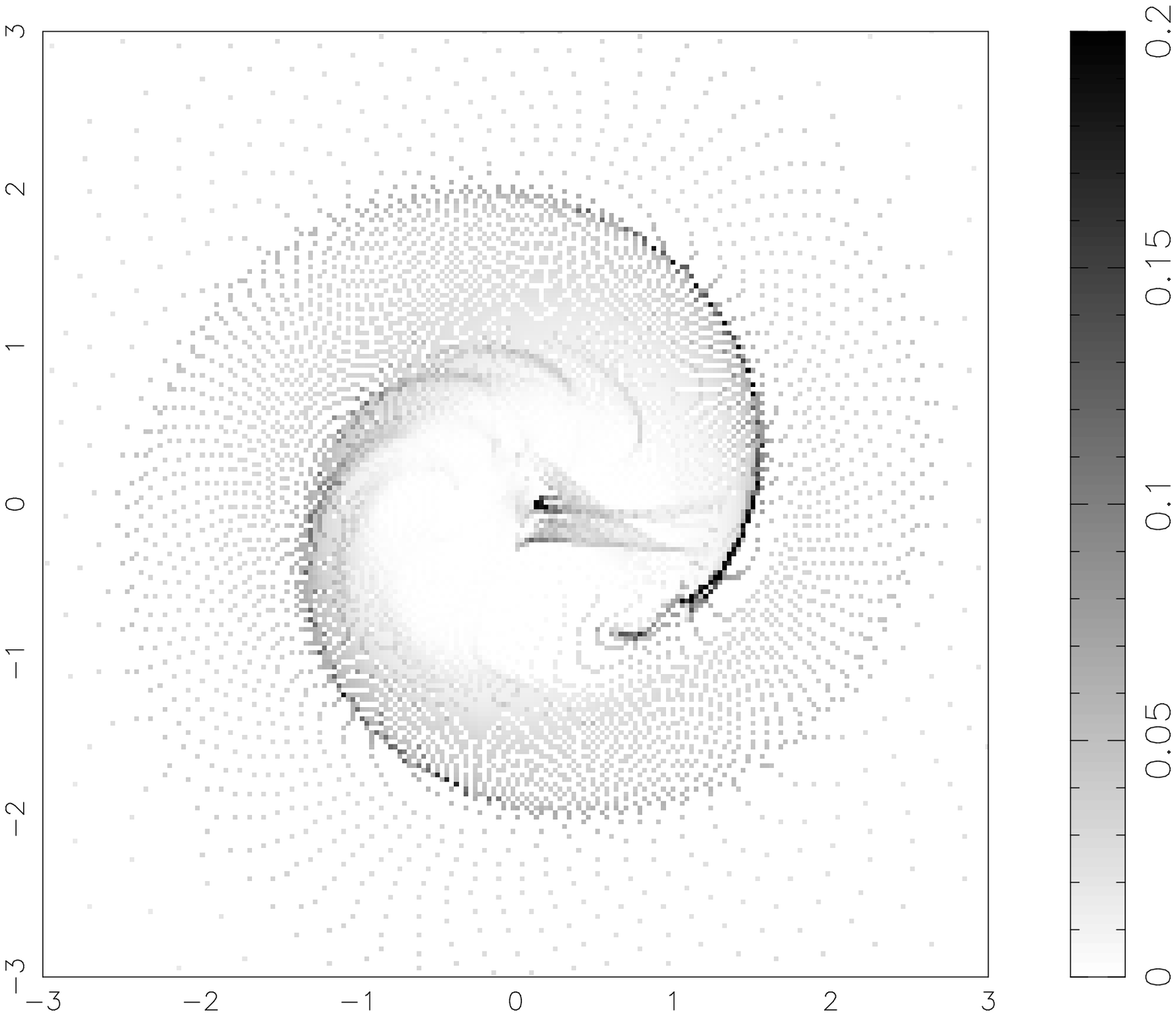}&
\epsfxsize=72mm
\epsfbox{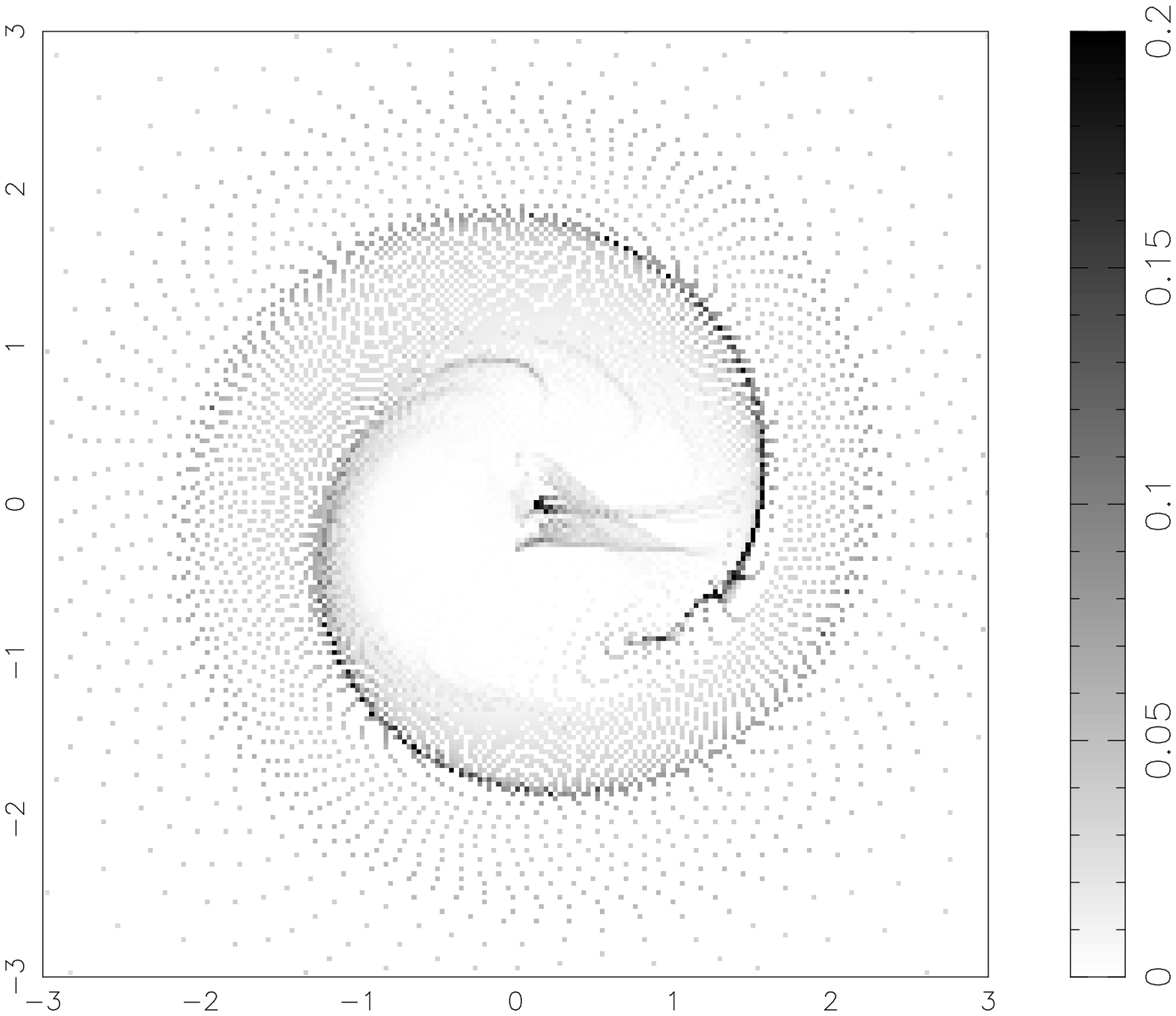}\\
\vspace{-5mm}\hspace{2.5cm}{\Large$\gamma=1.05$} & 
\vspace{-5mm}\hspace{2.5cm}{\Large$\gamma=1.01$} \\
\end{tabular*}
\end{center}
\caption[dopxy2d]{Doppler maps (hodograph) of 2D calculations in the orbital plane at 5 revolution periods. Density profiles in velocity coordinates ($V_x, V_y$) are shown.}
\label{fig:dopplerxy2d}
\end{figure*}

\begin{figure*}
\begin{center}
\begin{tabular*}{1.0\columnwidth}{p{0.5\columnwidth}p{0.5\columnwidth}}
\leavevmode
\epsfxsize=72mm
\epsfbox{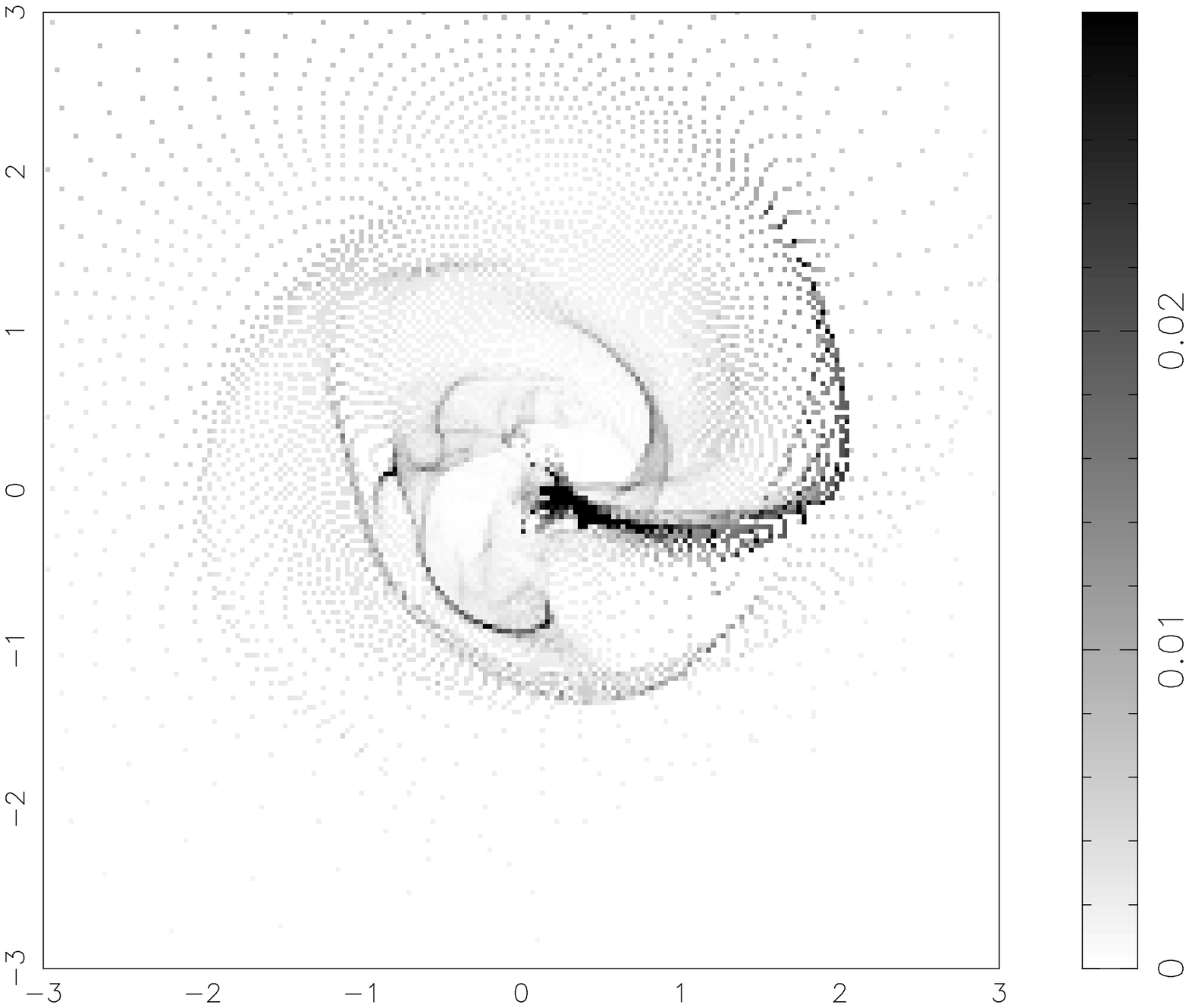}&
\epsfxsize=72mm
\epsfbox{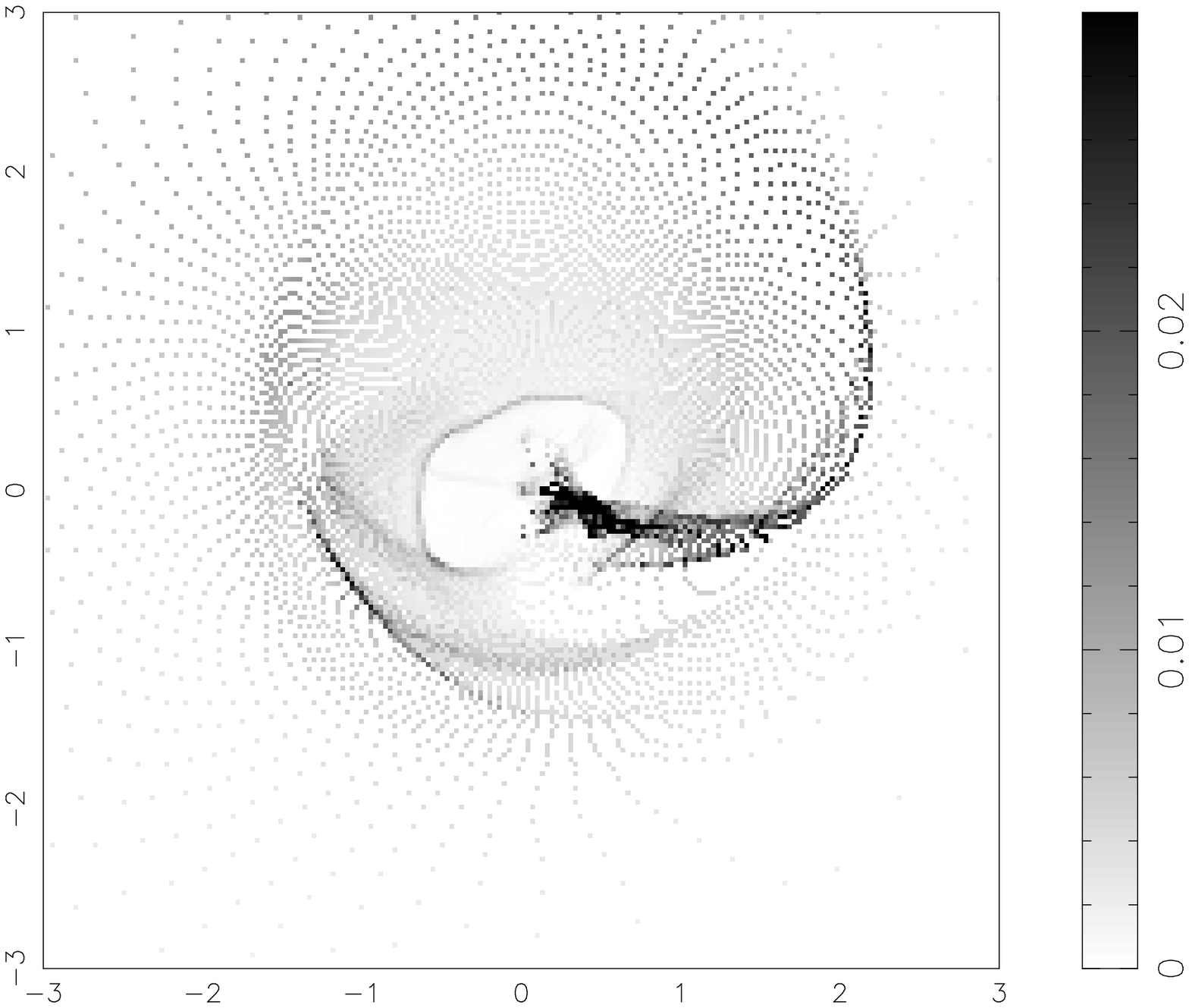}\\
\vspace{-5mm}\hspace{25mm}{\Large $\gamma=1.2$}&
\vspace{-5mm}\hspace{25mm}{\Large $\gamma=1.1$}\\
\epsfxsize=72mm
\epsfbox{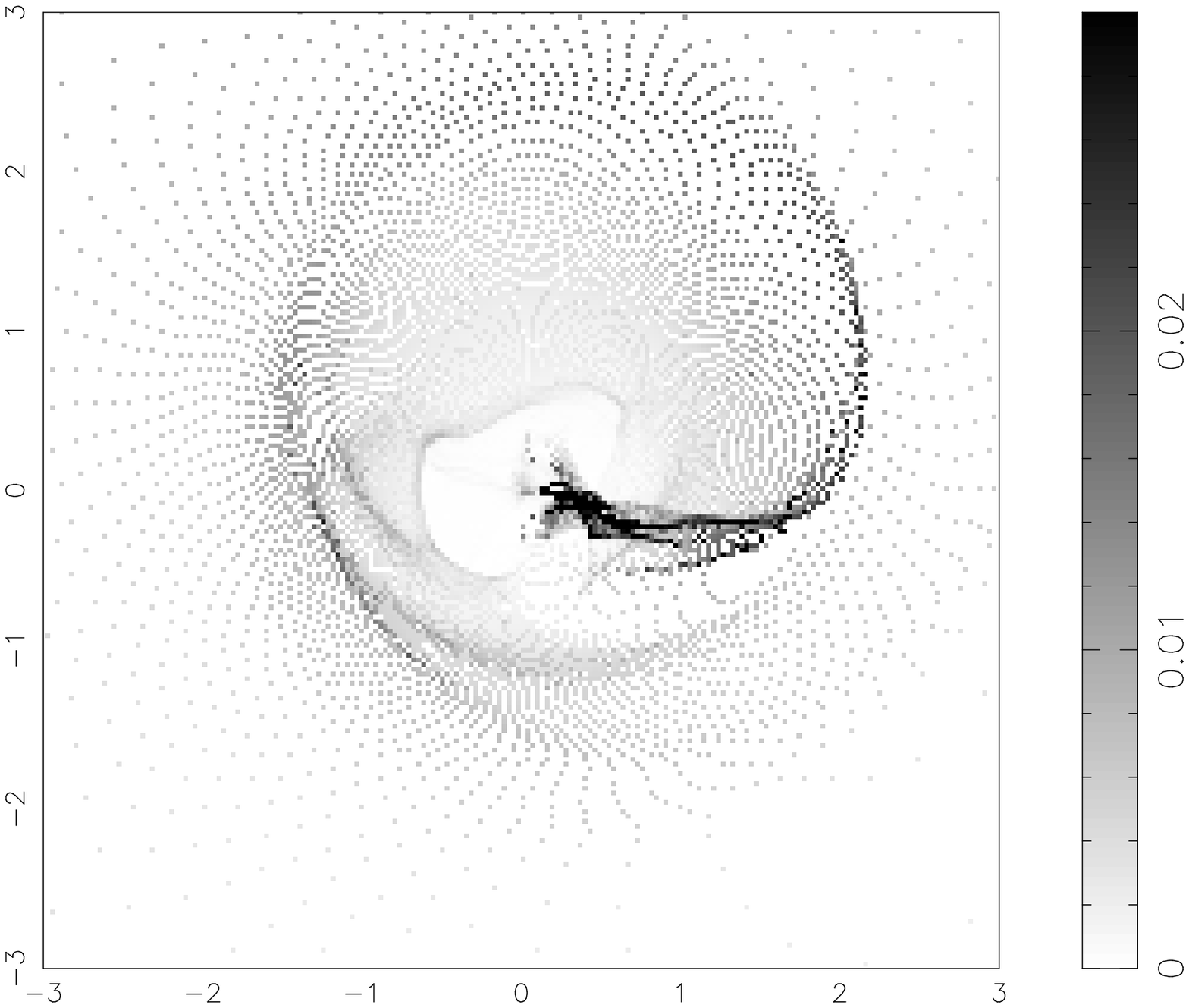}&
\epsfxsize=72mm
\epsfbox{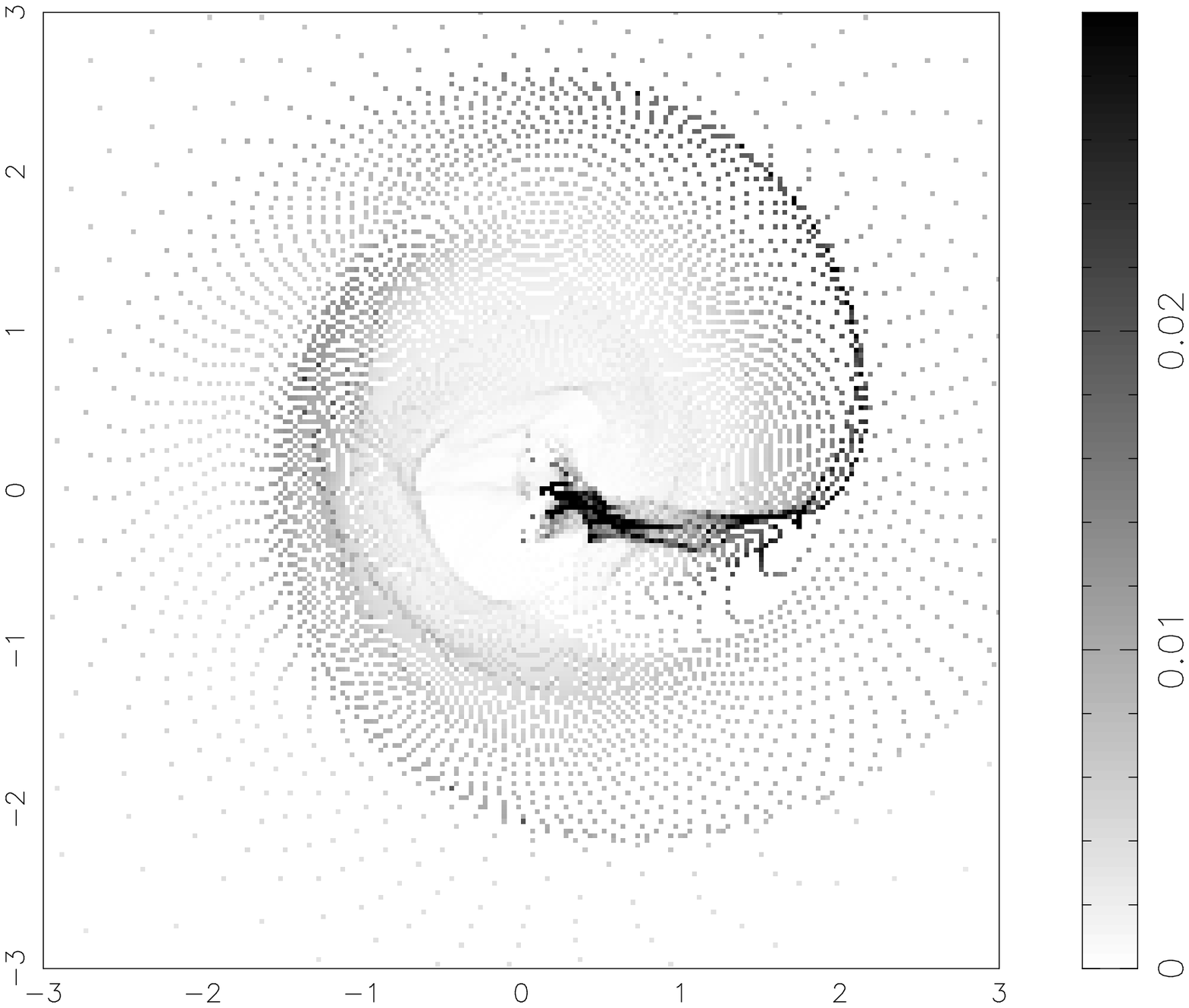}\\
\vspace{-5mm}\hspace{2.5cm}{\Large$\gamma=1.05$} & 
\vspace{-5mm}\hspace{2.5cm}{\Large$\gamma=1.01$} \\
\end{tabular*}
\end{center}
\caption[dopxy3d]{Doppler maps (hodograph) of 3D calculations in the orbital plane at 5 revolution periods. Density profiles in velocity coordinates ($V_x, V_y$) are shown.}
\label{fig:dopplerxy3d}
\end{figure*}

Figs. \ref{fig:dopplerxy2d} and \ref{fig:dopplerxy3d} show density distribution in Doppler coordinates ($V_x, V_y$) for 2D and 3D cases.
From these maps, we find the clear spiral structure in both 2D and 3D (and the stream from L1 point.)
We, therefore, conclude that there exists the spiral structure in real physical space.
Since the mass ratio of the binary in IP Peg is not 1 but 0.5, the direct comparison between Figs. 10 or 11 with the observation may be meaningless.
We calculated 2D flows for the mass ratio of 0.5 and compared with the observation (Matsuda et al. 1998) and obtained a reasonable agreement.
The comparison between 3D calculations and the observations is now under way.
 
\section{Conclusion}
We have performed 2D and 3D numerical simulations of accretion discs in close binary systems using SFS finite volume method.
We obtain the following conclusions:
\begin{enumerate}
\item In 2D calculations we observe clear spiral shocks, which winds more tightly for smaller $\gamma$. (See Fig. \ref{fig:dense2d}).
This result is consistent with previous works.
\item In 3D cases we confirm that discs are formed even for $\gamma \geq 1.1$ contrary to the claims by Lanzafame et al. (1992) and Bisikalo et al. (1998c) 
(see Fig. \ref{fig:isodense}), but this is consistent with the previous works (Sawada \& Matsuda 1992; Yukawa et al. 1997).
\item In 3D calculations we also find spiral shocks for all four cases, i.e. $\gamma=1.2, 1.1, 1.05$ and 1.01, as well as 2D cases.
\item However, the pitch angle of the spirals in 3D is not so markedly correlated with $\gamma$ as in 2D cases. (See Figs. \ref{fig:dense2d} and \ref{fig:densex-y}).
\item The stream from the L1 point does not smoothly penetrate into the accretion disc, of which fact contradicts with the claim by Bisikalo et al. (1998b, c). (See Fig. \ref{fig:isodense}).
\item Spiral Shocks are produced by the tidal interaction contrary to the claim by Bisikalo et al. (1998b, c). (See Fig. \ref{fig:11ni}).
\item Doppler maps for 2D and 3D are constructed.
Such maps can be compared with observations.
\end{enumerate}

\section*{Acknowledgments}
This work was supported in part by the Grant-in-Aid for the Ministry of Education, Science, and Culture of Japan (05640350, 10640231). The 2D calculations were mainly performed on Fujitsu VX/1R at the National Astronomical Observatory, Japan.
The 3D calculations were performed on NEC SX-4 at the information processing centre of Kobe University.

\appendix

\section[]{SFS scheme}
SFS scheme is one of Advection Upstream Splitting Method (AUSM) type schemes.
Original AUSM scheme invented by Liou \& Steffen (1993) is very simple, robust for strong shock and accurate, but shows overshoot at shock front.
Inspired by AUSM, various schemes including SFS have been proposed since then.
Although, various schemes have been developed independently, these schemes can be written in a common form.
They can be called as AUSM type schemes. 
 
Now we show the formulation of AUSM type scheme for three-dimensional Euler equation. (See Shima \& Jounouchi 1997).
Three-dimensional Euler equation can be written in the integral form as follows:
\begin{equation}
\int \tilde{Q} dv+\int \tilde{F} ds=0,
\end{equation} 
\begin{equation}
\tilde{Q} =\left(
\begin{array}{c}
\rho \\
\rho u \\
\rho v \\
\rho w \\
e
\end{array}\right),
\end{equation}
\begin{equation}
\tilde{F}=m\Psi+pN, \quad\Psi=\left(
\begin{array}{c}
1\\
u\\
v\\
w\\
h
\end{array}\right)
,\quad N=\left(
\begin{array}{c}
0\\
x_n\\
y_n \\
z_n \\
0
\end{array}\right),
\end{equation}
\begin{equation}
m=\rho V_n,\quad V_n=x_nu+y_nv+z_nw,
\end{equation}
where $\rho, u, v, w, e, p,$ and $h=(e+p)/\rho$ represent the density, the velocity in the x, y and z direction, the total energy per unit volume, the pressure and the total enthalpy, respectively. 
$x_n$, $y_n$ and $z_n$ show unit normals to the surface.
This form suggests that Euler flux can be divided into the advection term and the pressure term.
AUSM is derived based on the fact that the advection term and the pressure term can be upwinded separately.

AUSM type schemes can be written in the following form,
\begin{equation}
\tilde{F}=\frac{m+|m|}{2}\Psi_{+}+\frac{m-|m|}{2}\Psi_{-}+\tilde{p}N,
\end{equation}
where subscript $\pm$ show physical value at left (+) and right ($-$) side of a cell boundary, and $\tilde{p}$ does mixing of pressure using Mach number of left and right state which is defined by 
\begin{equation}
\tilde{p}=\beta_{+}p_{+}+\beta_{-}p_{-},
\end{equation}
\begin{equation}
\beta_{\pm}=\frac{1}{4}(2\mp{\cal M_{\pm}})({\cal M_{\pm}}\pm1)^2,\quad if \enskip|{\cal M_{\pm}}|\leq 1,
\end{equation}
and $\beta_{\pm}$ are smoothly switched to 1 or 0 for supersonic case.

By choosing different forms for the mass flux $m$ above, various AUSM type schemes have been proposed.
SFS uses variations of van Leer's Flux Vector Splitting (FVS) for its mass flux.
This is written in following form,
\begin{equation}
m=m_{+}+m_{-},
\end{equation}
\begin{equation}
m_{\pm}=\pm\frac{\rho_{\pm}c^2_{\pm}}{4\bar{c}}({\cal M_{\pm}}\pm 1)^2, \quad if\enskip |{\cal M_{\pm}}|\leq 1,
\end{equation}
\begin{equation}
{\cal M_{\pm}}=\frac{V_{n\pm}}{\bar{c}_{\pm}}, \quad \bar{c}_{\pm}=\frac{c^2_{\pm}}{\bar{c}}, \quad c^2_{\pm}=\gamma\frac{p_{\pm}}{\rho_{\pm}}, 
\end{equation}
where 
\begin{equation}
\bar{c}=\sqrt{\frac{c^2_{+}+c^2_{-}}{2}}.
\end{equation}
$\bar{c}$ is arithmetic average of sound speed and pure upwind side value or 0 of ${\cal M}_{\pm}$ will be used for supersonic case.
SFS has improved overshoot at a shock.

\bsp
\label{lastpage}

\begin{thebibliography}{99}
\bibitem{bi97a}
Bisikalo D.V., Boyarchuk A.A., Kuznetsov O.A., Chechetkin V.M., 1997a, Astron. Reports, 41, 786
\bibitem{bi97b}
Bisikalo D.V., Boyarchuk A.A., Kuznetsov O.A., Chechetkin V.M., 1997b, Astron. Reports, 41, 794
\bibitem{bi98a}
Bisikalo D.V., Boyarchuk A.A., Kuznetsov O.A., Khruzina T.S., Cherepashchuk A. M., Chechetkin V.M., 1998a, Astron. Reports, 42, 33
\bibitem{bi98b}
Bisikalo D.V., Boyarchuk A.A., Chechetkin V.M., Kuznetsov O.A., Molteni D., 1998b, MNRAS, in press
\bibitem{bi98c}
Bisikalo D.V., Boyarchuk A.A., Kuznetsov O.A., Chechetkin V.M., 1998c, Astron. Reports, in press
\bibitem{Ch90}
Chakrabarti S.K., 1990, ApJ, 362, 406
\bibitem{go97}
Godon P., 1997, ApJ, 480, 329
\bibitem{hi91}
Hirose M., Osaki Y., Mineshige S., 1991 PASJ, 43, 809
\bibitem{Jyo93}
Jyounouchi T., Kitagawa I., Sakashita, Yasuhara M., 1993, {\it Proceedings of 7th CFD Symposium} 
\bibitem{la92}
Lanzafame G., Belvedere G., Molteni D., 1992, MNRAS, 258, 152 
\bibitem{la93}
Lanzafame G., Belvedere G., Molteni D., 1993, MNRAS, 263, 839
\bibitem{la97}
Lanzafame G., Belvedere G., 1997, MNRAS, 284, 957 
\bibitem{la98}
Lanzafame G., Belvedere G., 1998, MNRAS, 295, 618
\bibitem{li90a}
Lin D.N.C., Papaloizou J.C.B., Savonije G.J., 1990a, ApJ, 364, 326
\bibitem{li90b}
Lin D.N.C., Papaloizou J.C.B., Savonije G.J., 1990b, ApJ, 365, 748
\bibitem{li93}
Liou M. S., Steffen C. J., 1993, J. Comp. Phys., 107, 23
\bibitem{lu93}
Lubow S.H., Pringle J.E., 1993, ApJ, 409, 360  
\bibitem{mar88}
Marsh T.R., Horne K., 1988, MNRAS, 235, 269
\bibitem{ma90}
Matsuda T., Sekino N., Shima E., Sawada K., Spruit H., 1990, A \& A, 235, 211
\bibitem{ma98}
Matsuda T., Makita M., Yukawa H., Boffin H.M.J., 1998, in Miyama S.M., Tomisaka K., Hanawa T., eds., Proceedings of Numerical Astrophysics 1998, Kluwer, in press  
\bibitem{mo91}
Molteni D., Belvedere G., Lanzafame G., 1991, MNRAS, 249, 748
\bibitem{ne98}
Neustroev V.V., Borisov N.V., 1998, A \& A, in press
\bibitem{na91}
Nagasawa M., Matsuda T., Kuwahara K., 1991, Numer. Astroph. in Japan, 2, 27
\bibitem{ro89}
Rozyczka M., Spruit H., 1989, in Meyer F., Duschl W.J., Frank J., Meyer-Hefmeister E., eds, Theory of Accretion Disks. Kluwer, Dordrecht, p.341
\bibitem{sav94}
Savonije G.J., Papaloizou J.C.B., Lin D.N.C., 1994, MNRAS, 268, 13
\bibitem{sa92}
Sawada K., Matsuda T., 1992, MNRAS, 255, s17
\bibitem{sa86a}
Sawada K., Matsuda T., Hachisu I., 1986a, MNRAS, 219, 75
\bibitem{sa86b}
Sawada K., Matsuda T., Hachisu I., 1986b, MNRAS, 221, 679
\bibitem{sa87}
Sawada K., Matsuda T., Inoue M., Hachisu I., 1987, MNRAS, 224, 307
\bibitem{sh73}
Shakura N.I., Sunyaev R.A., 1973, A\&A, 24, 337
\bibitem{Sh94}
Shima E., Jyounouchi T., 1994, {\it 25th Annual Meeting of Space and Aeronautical Society of Japan,} pp.36-37 
\bibitem{Sh97}
Shima E., Jyounouchi T., 1997, {\it NAL-SP34, Proceedings of 14th NAL symposium on Aircraft Computational Aerodynamics}, p.7
\bibitem{sp87}
Spruit H., 1987, A\&A, 184, 173
\bibitem{sp89}
Spruit H., 1989, in Meyer F., Duschl W.J., Frank J., Meyer-Hefmeister E., eds, Theory of Accretion Disks. Kluwer, Dordrecht, p.325
\bibitem{spe87}
Spruit H., Matsuda T., Inoue M., Sawada K., 1987, MNRAS, 229, 517
\bibitem{st97} 
Steeghs D., Harlaftis E.T., Horne K., 1997, MNRAS, 290, L28
\bibitem{ste96}
Steeghs D., Horne K., Marsh T.R., Donati J.F., 1996, MNRAS, 281, 626
\bibitem{yu97}
Yukawa H., Boffin H.M.J., Matsuda T., 1997, MNRAS, 292, 321
\end{thebibliography}
\end{document}